\documentclass[fleqn,usenatbib]{mnras}

\usepackage{newtxtext,newtxmath}
\usepackage[T1]{fontenc}

\DeclareRobustCommand{\VAN}[3]{#2}
\let\VANthebibliography\thebibliography
\def\thebibliography{\DeclareRobustCommand{\VAN}[3]{##3}\VANthebibliography}

\usepackage{graphicx}	% Including figure files
\usepackage{amsmath}	% Advanced maths commands
\usepackage{ae,aecompl}
\usepackage[utf8]{inputenc}
\usepackage{hyperref}
\hypersetup{
    colorlinks=true,
    linkcolor=blue,
    citecolor=blue,
}
\usepackage[normalem]{ulem}
\usepackage{multirow}
\usepackage{booktabs, blkarray}
\newcommand{\aperp}{\alpha_\perp}	
\newcommand{\apar}{\alpha_\parallel}
\newcommand{\fsig }{f \sigma_8}
\newcommand{\hmpc }{$h^{-1}$Mpc}
\newcommand{\ezmock}{\textsc{EZmock}}
\newcommand{\ezmocks}{\textsc{EZmocks}}
\newcommand{\nseries}{\textsc{Nseries}}
\newcommand{\patchy}{\textsc{MD-Patchy}}
\newcommand{\bigmd}{\textsc{BigMD}}
\newcommand{\lcdm}{$\Lambda$CDM}

\title[Void-galaxy correlation from eBOSS LRGs]{The Completed SDSS-IV extended Baryon Oscillation Spectroscopic Survey: geometry and growth from the anisotropic void-galaxy correlation function in the luminous red galaxy sample}

\author[S. Nadathur et al.]{\parbox{\textwidth}{
Seshadri Nadathur,$^{1}$\thanks{E-mail: seshadri.nadathur@port.ac.uk}
Alex Woodfinden,$^{2,3}$
Will J. Percival,$^{2,3,4}$
Marie Aubert,$^{5}$
Julian Bautista,$^{1}$
Kyle Dawson,$^{6}$
St\'ephanie Escoffier,$^{5}$
Sebastien Fromenteau,$^{7}$
H\'ector Gil-Mar\'in,$^{8,9}$
James Rich,$^{10}$
Ashley J. Ross,$^{11}$
Graziano Rossi,$^{12}$
Mariana Vargas Maga\~na,$^{13}$
Joel R. Brownstein,$^{6}$
Donald P. Schneider$^{14,15}$
}
\vspace*{4pt} \\
$^{1}$Institute of Cosmology and Gravitation, University of Portsmouth, Burnaby Road, Portsmouth, PO1 3FX, UK\\
$^{2}$Waterloo Centre for Astrophysics, University of Waterloo, 200 University Ave W, Waterloo, ON N2L 3G1, Canada\\
$^{3}$Department of Physics and Astronomy, University of Waterloo, 200 University Ave W, Waterloo, ON N2L 3G1, Canada\\
$^{4}$Perimeter Institute for Theoretical Physics, 31 Caroline St. North, Waterloo, ON N2L 2Y5, Canada\\
$^{5}${Aix Marseille Univ, CNRS/IN2P3, CPPM, Marseille, France}\\
$^{6}$Department of Physics and Astronomy, University of Utah, 115 S 1400 E, Salt Lake City, UT 84112, USA\\
$^{7}$Instituto de Ciencias F\'isicas, Universidad Nacional Aut\'onoma de M\'exico, Av. Universidad s/n, 62210 Cuernavaca, Mor., Mexico\\
$^{8}$Institut de Ciencies del Cosmos, Universitat de Barcelona, ICCUB, Mart\'i i Franqu\`es 1, E08028 Barcelona, Spain\\
$^{9}$Institut d’Estudis Espacials de Catalunya (IEEC), E08034 Barcelona, Spain\\
$^{10}$IRFU, CEA, Universit\'e Paris-Saclay, F-91191 Gif-sur-Yvette, France\\
$^{11}$Center for Cosmology and Astro-Particle Physics, Ohio State University, Columbus, Ohio, USA\\
$^{12}$Department of Physics and Astronomy, Sejong University, Seoul 143-747, Korea\\
$^{13}$Instituto de F\'isica, Universidad Nacional Aut\'onoma de M\'exico, Apdo. Postal 20-364, Ciudad de M\'exico, M\'exico\\
$^{14}$Department of Astronomy and Astrophysics, The Pennsylvania State University, University Park, PA 16802, USA\\ 
$^{15}$Institute for Gravitation and the Cosmos, The Pennsylvania State University, University Park, PA 16802, USA 
}

\date{Accepted XXX. Received YYY; in original form ZZZ}

\pubyear{2020}

\begin{document}
\label{firstpage}
\pagerange{\pageref{firstpage}--\pageref{lastpage}}
\maketitle

\begin{abstract}
We present an analysis of the anisotropic redshift-space void-galaxy correlation in configuration space using the Sloan Digital Sky Survey extended Baryon Oscillation Spectroscopic Survey (eBOSS) Data Release 16 luminous red galaxy (LRG) sample. 
This sample consists of LRGs between redshifts 0.6 and 1.0, combined with the high redshift $z>0.6$ tail of the Baryon Oscillation Spectroscopic Survey Data Release 12 CMASS sample. 
We use a reconstruction method to undo redshift-space distortion (RSD) effects from the galaxy field before applying a watershed void-finding algorithm to remove bias from the void selection. 
We then perform a joint fit to the multipole moments of the correlation function for the growth rate $f\sigma_8$ and the geometrical distance ratio $D_M/D_H$, finding $f\sigma_8(z_\rmn{eff})=0.356\pm0.079$ and $D_M/D_H(z_\rmn{eff})=0.868\pm0.017$ at the effective redshift $z_\rmn{eff}=0.69$ of the sample. 
The posterior parameter degeneracies are orthogonal to those from galaxy clustering analyses applied to the same data, and the constraint achieved on $D_M/D_H$ is significantly tighter. 
In combination with the consensus galaxy BAO and full-shape analyses of the same sample, we obtain $f\sigma_8=0.447\pm0.039$, $D_M/r_d=17.48\pm0.23$ and $D_H/r_d=20.10\pm0.34$. 
These values are in good agreement with the \lcdm{} model predictions and represent reductions in the uncertainties of $13\%$, $23\%$ and $28\%$ respectively compared to the combined results from galaxy clustering, or an overall reduction of 55\% in the allowed volume of parameter space. 
\end{abstract}

\begin{keywords}
cosmology: observations -- large-scale structure of Universe -- cosmological parameters
\end{keywords}

%%%%%%%%%%%%%%%%%%%%%%%%%%%%%%%%%%%%%%%%%%%%%%%%%%

%%%%%%%%%%%%%%%%% BODY OF PAPER %%%%%%%%%%%%%%%%%%

%%%%%%%%%%% INTRODUCTION %%%%%%%%%%%%
\section{Introduction}
\label{sec:intro}

The observed large-scale distribution of galaxies in the Universe encodes a wealth of physics. 
For cosmology, its importance lies in enabling measurement of the expansion history of the Universe and the growth of structure within it. 
These in turn allow precise tests of the components of the $\Lambda$ Cold Dark Matter (\lcdm) standard model, of gravity theories, the curvature of space and other important questions in modern physics \citep[e.g., see][]{eBOSS_Cosmology}. 

Optimally extracting this information from the data requires careful use of techniques designed to make measurements without systematic biases. 
The use of the baryon acoustic oscillation (BAO) standard ruler is one such technique; since the first BAO detections \citep{Percival:2001,Eisenstein:2005,Cole:2005} the method has matured and has been applied to several galaxy surveys at different redshifts \citep[e.g.,][]{Percival:2007,Beutler:2011hx,Padmanabhan:2012,Kazin:2014, Ross:2015,Alam:2017,duMasdesBourboux:2017}. 
The large datasets now available allow detection of the BAO feature both along and transverse to the line of sight direction and thus the use of the Alcock-Paczynski (AP) test \citep{Alcock:1979}, enabling geometric measurements of both $H(z)r_d$ and $D_M(z)/r_d$, where $H(z)$ is the Hubble expansion rate and $D_M(z)$ the transverse comoving distance to redshift $z$, and $r_d$ is the sound horizon at the baryon drag epoch. 

Redshift space distortions \citep[RSD;][]{Kaiser:1987} in the galaxy two-point statistics provide another source of information and have been widely studied \citep[e.g.,][]{Peacock:2001,Guzzo:2008,Blake:2012,Beutler:2012,Howlett:2015,Alam:2017} to measure both the growth rate and geometry. 
The degeneracy between RSD and the geometric AP effect at large scales \citep{Ballinger:1996} and non-linear behaviour at small scales limit the information that can be extracted in this way. 
This degeneracy is partially broken by the BAO feature at scales of $\sim100\;h^{-1}$Mpc in the correlation function, and further information can be added by pushing the theoretical modelling to smaller scales via effective field theory approaches \citep{Ivanov:2019a,Colas:2019,D'Amico:2019}. 

These approaches all use the full field of galaxies; i.e., they are based on the ensemble clustering of the galaxy distribution as a whole. 
However, additional information is available from regions of low density---known as cosmic voids---which remain unvirialized and do not suffer shell-crossing. 
In other words, voids represent the regions of the Universe where galaxy motions have deviated the least from their Zel'dovich predictions \citep{Zeldovich:1970}. 
As a consequence, RSD effects due to galaxy motions around voids can be modelled remarkably successfully down to much smaller scales using linear perturbation theory alone \citep{Paz:2013,Cai:2016a,Nadathur:2019a}. 
Thus, by selecting only voids we can extract cosmological information to smaller scales than from the population as a whole.
This fact has spurred many recent studies of the growth rate of structure using the void-galaxy correlation \citep{Hamaus:2016,Hamaus:2017a,Hawken:2017,Nadathur:2019c,Achitouv:2019,Hawken:2020,Aubert20a}, which match the wider recent interest in voids as novel cosmological probes \citep[e.g.][]{Pisani:2015,Sanchez:2016,Nadathur:2016b,Raghunathan:2020}.

However, the primary benefit of void-galaxy correlation analyses is in fact the precision with which they may be used in a version of the AP test \citep{Lavaux:2012,Hamaus:2016,Nadathur:2019c}. 
Assuming statistical isotropy of the Universe, the real-space void-galaxy correlation function measured for a large enough sample of voids should show spherical symmetry. 
This symmetry is broken by RSD; but as discussed above, these effects can now be very accurately modelled. 
This allows one to isolate possible additional anisotropies introduced by the conversion of measured galaxy redshifts to distances using an assumed fiducial model that differs from the true cosmology. 
While the absolute sizes of voids are not predicted by fundamental theory, this test of anisotropy provides constraints on the dimensionless ratio $D_M/D_H(z)$, where the `Hubble distance' $D_H\equiv c/H(z)$, with $c$ the speed of light. 
The anisotropies introduced in the void-galaxy correlation by RSD and AP effects are not strongly degenerate and may be easily separated from each other \citep{Nadathur:2019c}, and the statistical precision with which $D_M/D_H$ can been measured using voids far exceeds that obtained from BAO \citep{Hamaus:2017a,Nadathur:2019c}. 
Systematic errors in this measurement were first quantified by \citet{Nadathur:2019c}; we provide a more exhaustive analysis in this work.

Equally importantly, the errors on cosmological distance and growth rate parameters obtained from such a void-galaxy measurement are not strongly correlated with those obtained from the combination of standard BAO and RSD analyses of the galaxy two-point statistics \citep{Nadathur:2019c,Nadathur:2020a}. 
This means that information from the different techniques applied to the same survey data may be combined to provide consensus results yielding a large reduction in measurement uncertainties. 
As a consequence, \citet{Nadathur:2020a} showed that adding void-galaxy measurements from a single redshift bin of the Baryon Oscillation Spectroscopic Survey (BOSS) greatly improved the constraints from large-scale structure on dark energy and curvature.

In this paper we present an analysis of the void-galaxy correlation measured in the final Data Release 16 (DR16) luminous red galaxy (LRG) sample of the extended Baryon Oscillation Spectroscopic Survey \citep[eBOSS;][]{Dawson:2013}, part of the fourth generation of the Sloan Digital Sky Survey \citep[SDSS;][]{Blanton:2004aa}. 
The DR16 release includes all eBOSS observations. The eBOSS LRG catalogue is combined with the $z>0.6$ high-redshift tail of the BOSS DR12 CMASS sample to form a combined catalogue of 377,458 well-understood, high-bias tracers covering the redshift range $0.6<z<1.0$, described in \citet{Ross20a}. 
We use this composite sample, whose galaxy properties closely match the lower redshift CMASS catalogue previously used for a similar void-galaxy analysis by \citet{Nadathur:2019c}. 
Traditional BAO and RSD analyses of the galaxy clustering for the same sample, in Fourier and configuration spaces, have been described by \citet{gil-marin20a, LRG_corr}, with systematic errors quantified in \citet{rossi20a}. 
Previous analyses for the LRG sample in SDSS Data Release 14 \citep[DR14;][]{Abolfathi:2018}, obtained using the first two years of data, were presented in \citet{Bautista:2018} and \citet{Icaza-Lizaola:2020}.

In addition to the LRGs, eBOSS also explores large-scale structure at higher redshifts out to $z<2.2$ using emission line galaxies (ELGs) and quasars as additional tracers of the density field that are not considered in this work. 
The ELG sample suffers from significant angular fluctuations because it was selected from imaging data with anisotropic properties \citep{raichoor20a}, and BAO \& RSD analyses \citep{tamone20a,demattia20a} have had to carefully correct for these effects. 
The quasar sample pushes to higher redshifts with a low-density sampling \citep{Ross20a,lyke20a}, and has also been used to make BAO \& RSD measurements \citep{hou20a,neveux20a}, also using mock catalogues to determine errors \citep{Smith:2020}. 
At redshifts $z>2.1$, measurements of BAO in the Lyman-$\alpha$ forest of a high redshift quasar sample are given in \citet{duMasdesBourboux:2020}. 
The cosmological interpretation of all of these BAO and RSD results from eBOSS samples was presented in \citet{eBOSS_Cosmology}. 
Finally, \citet{Aubert20a} presented related measurements of the growth rate around voids in the LRG, ELG and quasar samples using a somewhat different model to that used in this paper.

The layout of this paper is as follows. In Section~\ref{sec:data} we describe the characteristics of the LRG data sample and the various mock catalogues used in this paper. 
In Section~\ref{sec:measurement} we describe the methods used to obtain void catalogues from the galaxy data and to estimate the void-galaxy correlation, while Section~\ref{sec:methods} lays out the details of the model and the fitting procedure. 
We present results of the void-galaxy fit in isolation in Section~\ref{sec:results}, and perform a thorough check of possible systematic error contamination in Section~\ref{sec:systematics}. 
In Section~\ref{sec:consensus} we then describe how the void-galaxy measurements are combined with those from galaxy clustering and present the consensus results from the eBOSS+CMASS sample after performing this combination. 
We conclude in Section~\ref{sec:conclusions}. 
Additional material on a comparison of alternative void-galaxy models and the associated systematic errors is presented in Appendix~\ref{sec:appendixA}.

%%%%%% DATA %%%%%%
\section{Data \& Mocks}
\label{sec:data}

%%%%%% FIGURE %%%%%%
\begin{figure}
    \centering
    \includegraphics[width=0.95\columnwidth]{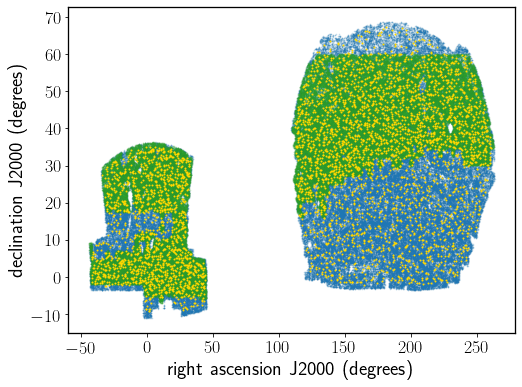}
    \caption{The footprint of the joint eBOSS+CMASS luminous red galaxy sample, showing the north and south galactic caps (NGC and SGC respectively). 
    Blue points show the distribution of CMASS galaxies from the BOSS DR12 release, which extend to a maximum redshift of $z\sim0.8$. The overlapping green points show the eBOSS LRGs, which cover $0.6<z<1.0$. 
    The yellow points show the locations of void centres, and are more heavily concentrated in the region common to the eBOSS and CMASS footprints.} 
    \label{fig:footprint}
\end{figure}

%%%%%% FIGURE %%%%%%
\begin{figure}
    \centering
    \includegraphics[width=0.95\columnwidth]{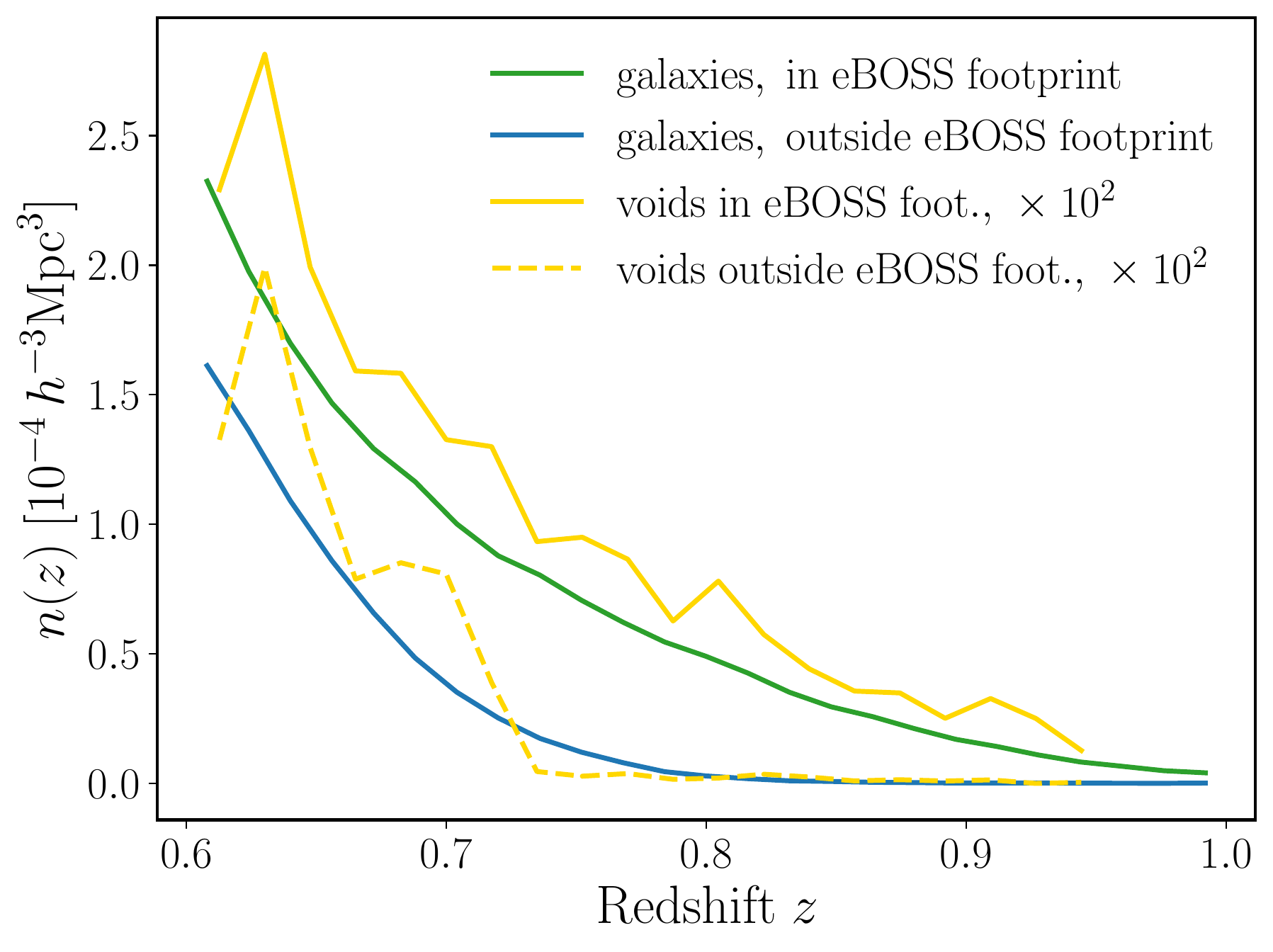}
    \caption{The number density of galaxies in the joint eBOSS+CMASS sample lying within the eBOSS footprint (green) and outside it (blue) as a function of redshift. 
    Differences in the number density for these two categories between the north and south galactic caps are very small so we show here the mean number density over the combined volume. 
    The yellow solid and dashed lines show the number density of voids inside and outside the eBOSS footprint respectively, multiplied by a factor of 100.
    } 
    \label{fig:nz_sample}
\end{figure}

\subsection{LRG sample}

eBOSS LRG spectra were obtained using the BOSS spectrographs \citep{Smee:2013} mounted on the 2.5-meter telescope \citep{Gunn:2006} at the Apache Point Observatory in New Mexico. 
The target sample was selected \citep{Prakash:2016} from SDSS DR13 photometry \citep{Albareti:2017}, with additional infrared information from the WISE satellite \citep{Lang:2016}. 
Over 7500\,deg$^{2}$, 60\,deg$^{-2}$ LRG targets were selected, of which 50\,deg$^{-2}$ were spectroscopically observed. 
The redshift of each LRG was estimated from its spectrum using the \textsc{redrock} algorithm. 
This uses templates derived from principal component analysis of SDSS data to classify spectra, which is followed by a redshift refinement procedure that uses stellar population models for galaxies. 
On average, 96.5 per cent of spectra yield a confident redshift estimate (details can be found in \citealt{Ross20a}). 

The creation of the LRG large-scale structure catalogue, a selection of the observed galaxies together with weights to correct for various effects, and a matched random catalogue that Monte-Carlo samples the observed region matching the galaxy completeness is presented in \citet{Ross20a}. 
The random catalog contains an unclustered set of spatial locations with the same radial and angular selection functions as the data. 
We use a random sample with 50 times more points than there are galaxies, to minimize the shot noise contribution from this catalogue. 
Redshifts for the randoms were sampled from galaxy redshifts in order to ensure that the radial distributions match.  

The galaxy and random catalogues are matched using a series of masks that eliminate regions with bad photometric properties, targets that collide with quasar spectra, which are selected at higher priority than co-observed LRGs, and the centerpost region of the plates where it is physically impossible to put a fiber. 
All masks combined cover 17 per cent of the initial footprint. 
About 4 per cent of the LRG targets were not observed due to fiber collisions, when a group of two or more galaxies are closer than 62$^{\prime\prime}$ so that because of hardware limitations, they cannot all receive a fiber \citep{Mohammad:2020}. 
For this analysis, the small-scale clustering in these high-density regions is not important, and we simply upweight the observed galaxy in a close-pair where one galaxy was missed by $w_{\rm cp}$ to correct for fiber collisions. 
To correct for the $3.5$ per cent of spectra that do not result in a reliable redshift estimate, we include a weight $w_{\rm noz}$, calculated as a function of position of the spectra on the detector and the signal-to-noise of that set of observation, to boost galaxies with good redshifts such that this weighted sample is an unbiased sampling of the full population. 
Systematic variations in the photometric data used for target selection are removed from the large-scale structure sample by weighting galaxies by weights $w_{\rm sys}$. 
These weights are computed with a multi-linear regression on the observed relations between the angular over-densities of galaxies versus stellar density, seeing and galactic extinction. 
FKP weights $w_{\rm FKP}$ that balance the signal given the variation in density across the sample \citep{Feldman:1994} are also included. 
For the reconstruction step described in Section~\ref{sec:recon} we use the combined weight defined as $w_\mathrm{tot} = w_{\rm noz}w_{\rm cp}w_{\rm sys} w_{\rm FKP}$.

The eBOSS sample of LRGs overlaps in area and redshift range with the high-redshift tail of the BOSS CMASS sample. 
To improve the signal, we combine the eBOSS LRG sample with all the $z > 0.6$ BOSS CMASS galaxies including non-overlapping areas. 
Overall, the high redshift BOSS CMASS galaxies represent 54 per cent of the sample used in this work. 
In the following, we refer to the combined sample as the eBOSS+CMASS sample, or where there is no risk of confusion, simply as the LRG sample. 
The angular footprints of the eBOSS and CMASS samples are shown in Fig.~\ref{fig:footprint}. 
The number densities of galaxies in the combined sample are thus very different in the regions only observed by CMASS and in the overlap of the CMASS and eBOSS footprints, as presented in Fig. \ref{fig:nz_sample}.

\subsection{Mocks}
In this work we employ several types of mock galaxy catalogues in order to estimate the covariance matrix of our data measurement, quantify the magnitude of possible systematic errors, and to calibrate fitting functions used in the theoretical modelling. 
These are described below.

\subsubsection{\emph{\ezmocks}}
The \ezmock{} catalogues are a set of 1000 independent mock galaxy catalogues created to closely mimic the clustering properties of the eBOSS+CMASS LRG sample. 
The \ezmock{} algorithm is based on a fast approximate Zeldovich method \citep{Chuang:2015}, together with deterministic and stochastic bias relations, a PDF mapping scheme and additional corrections to account for RSD, as described in detail by \citet{zhao20a}. 
The mocks are created on a lightcone, by combining the output from 4 and 5 different redshift snapshots for the CMASS and eBOSS LRG samples respectively, and are trimmed to match the survey volume, with the survey veto mask, radial selection, angular systematics and fibre collisions imprinted. 
The mocks are constructed using a flat \lcdm{} cosmology, with $\Omega_m=0.307$, $\Omega_b=0.0482$, $h=0.6777$, $\sigma_8=0.8225$ and $n_s=0.96$.

We use the \ezmocks{} to generate void random catalogues for use in the correlation estimator described in Section \ref{sec:measurement}, to estimate the covariance matrix of our void-galaxy measurement, and to determine the cross-covariance in parameter estimates between the void-galaxy method and galaxy clustering in Fourier \citep{gil-marin20a} and configuration space \citep{LRG_corr} applied to the same LRG sample. 

\subsubsection{\emph{\nseries} mocks}
The \nseries{} mocks are set of full $N$-body simulations, generated using a flat \lcdm{} cosmology with $\Omega_m=0.286$, $\Omega_b=0.0470$, $h=0.70$, $\sigma_8=0.82$ and $n_s=0.96$, with box side $2.6\;h^{-1}$Gpc, $2048^3$ particles per box and a mass resolution of $1.5\times10^{11}\;M_\odot/h$. 
They are populated with mock galaxies using a Halo Occupation Distribution (HOD) model chosen to reproduce the clustering properties of the BOSS DR12 CMASS galaxy sample. 
Seven independent cubic boxes were generated and from each box, 12 mock cut-sky catalogues matching the CMASS NGC sample were created using different projections and cuts, to create a total of 84 pseudo-independent mock catalogues. 
These mocks match the survey geometry, galaxy $n(z)$ and clustering of the BOSS CMASS sample, rather than those of the eBOSS+CMASS LRGs, and have a lower effective redshift, $z_\mathrm{eff}=0.55$. 
However, the large available volume, $84\,\times\,3.67\;\mathrm{Gpc}^3$, and the accurate RSD makes them ideal for testing potential modelling systematics. 

\subsubsection{\emph{\patchy} mocks}
The \patchy{} mocks are a suite of fast approximate mock galaxy catalogues on the lightcone created using the \textsc{Patchy} algorithm \citep{Kitaura-PATCHY:2014,Kitaura-DR12mocks:2016} and designed to match the BOSS DR12 CMASS sample. 
We use 1000 of these mocks in the NGC region in order to estimate the covariance matrix of the measurement performed on the \nseries{} mocks. 

\subsubsection{\emph{\textsc{Big MultiDark}} mocks}
The Big MultiDark simulation \citep{Klypin:2016} is a full $N$-body simulation which evolved $3840^3$ dark matter particles in a box of side $2.5\;h^{-1}$Gpc using the same cosmology as that of the \ezmocks. 
The particle mass resolution of the simulation is $2.359\times10^{10}\;M_{\odot}/h$. We use the halo catalogue from this simulation at redshift $z=0.70$, populated with an HOD matching that of \citet{Tinker:2017} to model the eBOSS+CMASS LRG sample, and the halo catalogue at redshift $z=0.52$, populated with the HOD from \citet{Manera:2013}, to model the CMASS sample. 
From these two boxes, we extract one cut-sky mock catalogue each matching the geometry and selection function of the respective samples, which we refer to as the \bigmd{} mocks.

The purpose of the \bigmd{} mocks is to make use of the full $N$-body dark matter particle information in order to estimate the void matter density profile $\delta(r)$ (i.e., the void-matter cross-correlation monopole) and the velocity dispersion profiles of galaxies around void centres, $\sigma_{v_\parallel}(r)$. 
These functions are used to calibrate the theory predictions for the void-galaxy correlation according to the method of \citet{Nadathur:2019c}, described in more detail in Section~\ref{sec:methods} below. 

\subsection{Reference cosmology}
\label{sec:refcosmo}
When analysing the DR16 data and the \ezmocks, we adopt a reference fiducial cosmology with parameters $\Omega_m=0.310$, $\Omega_\Lambda=0.69$, $h=0.676$ and zero curvature, in order to convert galaxy redshifts to distances. 
This reference cosmology choice is motivated by the CMB results from \citet{Planck:2018params} and is the baseline cosmology adopted for all eBOSS analyses. 
It is  close to, but not the same as, the true cosmology of the \ezmocks, which have $\Omega_m = 0.307$. 
In Section~\ref{sec:model_syst} we test the sensitivity of the parameter inference to this arbitrary choice of fiducial model by analysing the \nseries{} mocks in a number of different cosmological models, including their own true cosmology. 

%%%%%% FIGURE %%%%%%
\begin{figure}
    \centering
    \includegraphics[width=\columnwidth]{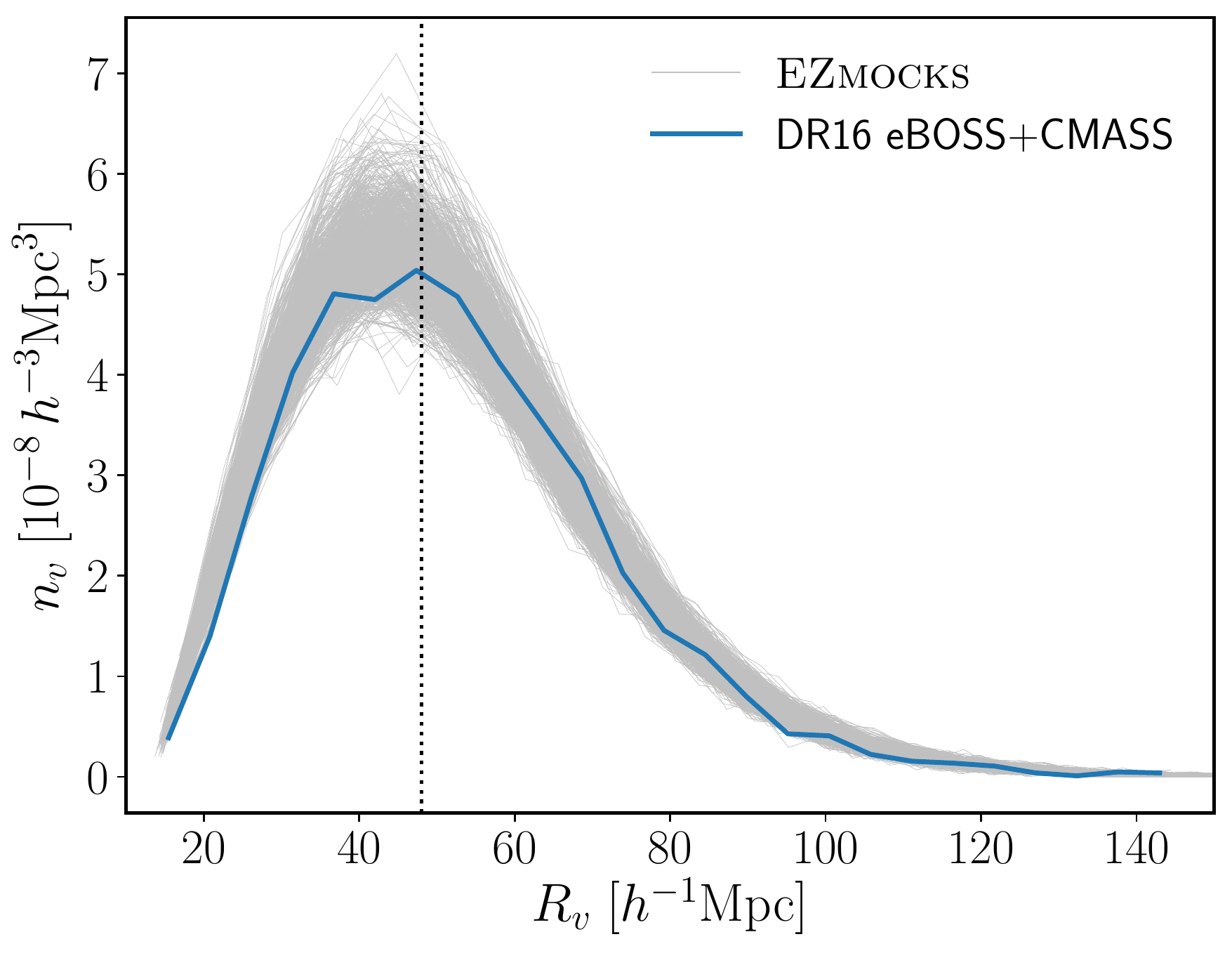}
    \caption{The void size function, i.e. the number density of voids in the catalogue as a function of their size, expressed in terms of an effective spherical radius $R_v$, for each of the 1000 \ezmocks{} and the DR16 LRG data. 
    This plot is for void catalogues obtained using the fiducial value $\beta=0.35$ for RSD removal via the reconstruction step before void-finding; the total number of voids in the catalogue changes by up to $\pm2\%$ as $\beta$ is varied over the range $[0.15, 0.55]$. 
    The dashed vertical line indicates the median void size, $R_v=49\;h^{-1}$Mpc, which is used as a minimum size cut in selecting the final void sample for analysis.
    } 
    \label{fig:void_sizes}
\end{figure}

%%%%%% FIGURE %%%%%%
\begin{figure*}
    \centering
    \includegraphics[width=0.95\linewidth]{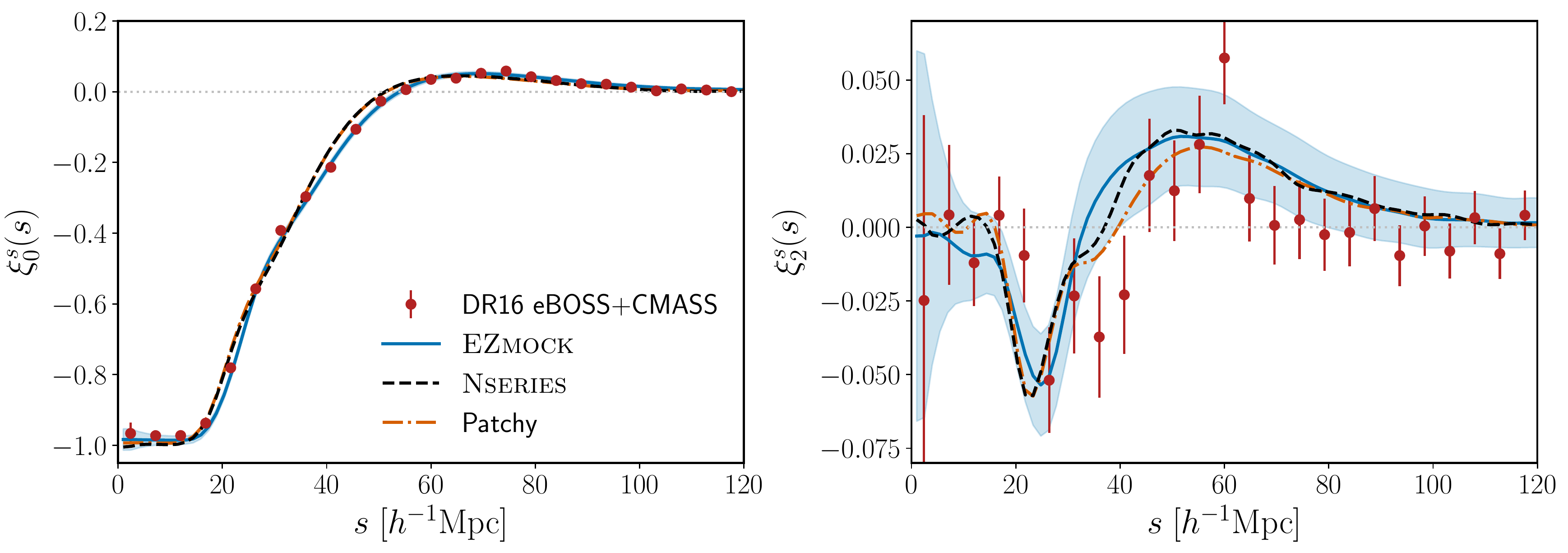}
    \caption{Monopole (left) and quadrupole (right) moments of the void-galaxy correlation function measured in the data sample and the mocks using the methods described in Section~\ref{sec:measurement}. 
    The solid blue line shows the mean of the 1000 \ezmock{} measurements and the shaded blue region indicates the standard deviation around this, which is an estimate of the error on a single realisation. 
    Points are for the DR16 LRG data, with error bars derived from the \ezmocks. 
    The dashed and dot-dashed lines show the means of the 84 \nseries{} and 1000 \patchy{} mocks respectively at a lower effective redshift. 
    \nseries{} and \patchy{} differ from the \ezmocks{} and data in galaxy $n(z)$ and bias, which changes the mean void size as visible in the monopole.
    All multipole measurements have a dependence on $\beta$; the DR16 and \ezmock{} data is shown for the fiducial $\beta^\mathrm{fid}=0.35$ for this sample while the \nseries{} and \patchy{} measurements are for $\beta^\mathrm{fid}=0.40$ appropriate for the lower redshift BOSS CMASS sample. 
    } 
    \label{fig:data_vs_mocks}
\end{figure*}

%%%%%% MEASUREMENT %%%%%%
\section{Measurement}
\label{sec:measurement}
In the following we describe the methods used to measure the void-galaxy correlation in the data and mock samples.

\subsection{Reconstruction and RSD removal}
\label{sec:recon}
As pointed out in several previous works \citep{Chuang:2017,Nadathur:2019a,Nadathur:2019b, Nadathur:2019c}, all models of the void-galaxy correlation $\xi_\mathrm{vg}$ assume that the anisotropy introduced in in redshift space can be described in terms of the peculiar velocities of the \emph{galaxies alone}. 
This is equivalent to the assumption that there are no RSD effects in the distribution of void positions, that the galaxy peculiar velocity field around void centres is spherically symmetric in real space (i.e. that the relative galaxy outflow velocity from voids, $\mathbf{v}_g(\mathbf{r})$, depends on the magnitude $r=|\mathbf{r}|$ of the void-galaxy separation vector only), and that the cross-correlation of voids with real-space galaxy positions is similarly spherically symmetric. 
Both the latter two assumptions are violated if the void-finding algorithm has any directional dependence leading to a lack of spherical symmetry for the selected sample.
This will be the case for any void-finder applied directly to redshift-space galaxy positions, as galaxy RSD will mean regions with large line-of-sight outflow velocities appear to have lower densities in redshift space, and thus are more likely to be selected as voids.

In fact, \citet{Nadathur:2019b} showed with reference to simulation results that \emph{none} of these assumptions are satisfied when the void sample is identified using the galaxy positions in redshift space (for a related result for the Lyman-$\alpha$ and a general theorem, see \citealt{Seljak:2012}; \citealt{Correa:2020} further discuss these issues for voids). 
Therefore, in order to ensure the validity of the theoretical modelling, we apply an RSD-removal algorithm to the galaxy data to recover the (approximate) real-space galaxy positions before performing the void-finding step. 
The RSD-removal procedure follows that outlined by \citet{Nadathur:2019b} and \citet{Nadathur:2019c}: we use the iterative reconstruction algorithm of \citet{Burden:2014,Burden:2015}, implemented in the public \texttt{REVOLVER} code\footnote{\url{https://github.com/seshnadathur/Revolver}} to solve the Zeldovich equation in redshift space for the Lagrangian displacement field $\mathbf{\Psi}$ on a $512^3$ grid, and shift individual galaxy positions by $-\mathbf{\Psi}_\mathrm{RSD}$, where $\mathbf{\Psi}_\mathrm{RSD} = -f\left(\mathbf{\Psi}\cdot\hat{\mathbf{r}}\right)\hat{\mathbf{r}}$, with $f$ the growth rate of structure. 
The reconstruction algorithm here is the same as that used for BAO reconstruction in the accompanying eBOSS clustering analyses \citep{LRG_corr,gil-marin20a,demattia20a,raichoor20a}, with the only difference being that for BAO reconstruction galaxies are shifted by the value of the full displacement field, $-\mathbf{\Psi}$. This difference is because we do not wish to identify voids in the linearly-evolved galaxy field, but only in that with large-scale RSD effects removed.

The implementation of the reconstruction algorithm for RSD removal depends on the value of $\beta=f/b$, where $b$ is the effective linear galaxy bias of the sample. 
As a result, all void catalogues, void-galaxy correlation data vectors and covariance matrices inherit this $\beta$-dependence. This is accounted for during parameter inference as described in Section \ref{sec:methods} below. The reconstruction also requires the specification of a smoothing scale, $R_s$. 
For the DR16 data and the \ezmocks, we use $R_s=15\;h^{-1}$Mpc as for the BAO analyses \citep{LRG_corr,gil-marin20a}. 
For the \nseries{} and \patchy{} mocks, as the mean galaxy number density is significantly larger, we use the smaller value $R_s=10\;h^{-1}$Mpc previously determined to be optimal for that case \citep{Nadathur:2019b}. 

Note that this procedure is not expected to accurately remove all small-scale RSD features from the galaxy field. 
However from tests on the \nseries{} mocks we find that the RSD removal recovers the true real-space galaxy power spectrum to within $2\%$ accuracy on scales $k\lesssim0.2\,h\rmn{Mpc}^{-1}$. 
This is sufficient for the purpose of recovering the real-space void centre positions and ensuring the validity of the model described in Section~\ref{sec:model} below.

\subsection{Void catalogue creation}
\label{sec:voids}
Void-finding is performed on the post-reconstruction RSD-removed galaxy samples using a modified version of the \texttt{REVOLVER} code \citep{Nadathur:2019c}. 
\texttt{REVOLVER} implements a watershed void-finding algorithm adapted from the \texttt{ZOBOV} (ZOnes Bordering On Voidness) code \citep{Neyrinck:2008}. 
This algorithm works by estimating the local galaxy density field from the discrete galaxy distribution by performing a Voronoi tessellation of the survey volume. 
Each Voronoi cell corresponds to the region of space closer to an individual galaxy than to any other. 
Thus the inverse volume of the Voronoi cells, normalized by the mean volume of all cells, provides an estimate of the local galaxy density in units of the mean. 
Variation of the survey selection function with redshift, $n(z)$, is accounted for in the normalization by the mean through using redshift weights such that the galaxy density field is always expressed in units of the local mean density at that redshift. 
Observational systematics are accounted for using the systematic weights $w_\mathrm{sys,tot}=w_\mathrm{noz}\cdot w_\mathrm{cp}\cdot w_\mathrm{sys}$. 
FKP weights are designed to optimise the power spectrum estimation but are not helpful for void-finding, which depends on galaxy density, so are omitted here.
To ensure the tessellation is contained within the surveyed region, \texttt{REVOLVER} places a thin, high-density shell of buffer particles around the survey volume and within holes in the mask to terminate the tessellation. 
Details of this procedure are provided in \citet{Nadathur:2014a} and \citet{Nadathur:2016a}. 

From the galaxy density field thus reconstructed, \texttt{REVOLVER} identifies local minima of the density as the sites of voids, whose extents are determined by the watershed basin of the density field around the minima \citep{Neyrinck:2008} with no merging of neighbouring zones. 
The adjacency information of the tessellation field is used to remove any Voronoi cells adjacent to one of the buffer particles used to terminate the tessellation, so these edge regions are never included in voids. 
Void centres are determined as the centre of the largest sphere completely empty of galaxies that can be inscribed in the void; this is also the circumcentre of the positions of the galaxies with the four largest Voronoi cells in each zone, and is the most robust estimate of the true location of the minimum of the total matter density in the void \citep{Nadathur:2015b}. 
Void sizes are characterised by an effective radius, $R_v$, corresponding to the radius of a sphere of equal volume to the void (although individual voids are in general not spherical). 

While the publicly released version of \texttt{REVOLVER} can be directly applied to the \nseries{} mocks, its use for the composite eBOSS+CMASS sample---or for the \ezmock{} and \bigmd{} mock samples created to match it---requires several important modifications. 
The first of these is because of the complex survey geometry caused by the different eBOSS and CMASS survey footprints (Fig.~\ref{fig:footprint}). 
The eBOSS galaxies cover a smaller sky area but extend to redshifts $z=1$, whereas there are almost no CMASS galaxies at $z>0.75$. 
We account for this difference by modifying the placement of the shell of buffer particles to correctly enclose the composite survey volume. 

A second important effect arises because, as shown in Fig. \ref{fig:nz_sample}, the selection function for the combined sample is not uniform across the sky. 
Within the region common to both CMASS and eBOSS footprints, $n(z)$ at $z=0.6$ is a factor of $\sim1.5$ larger than that in the region outside the eBOSS footprint, and this difference increases significantly with redshift. We estimate the mean galaxy density as a function of redshift separately for galaxies inside and outside the eBOSS footprint, denoted $n_\mathrm{in}(z)$ and $n_\mathrm{out}(z)$ respectively. The redshift weights applied by \texttt{REVOLVER} correct for this variation of the local mean density; we therefore modify the algorithm to determine the redshift weights to apply using either $n_\mathrm{in}(z)$ or $n_\mathrm{out}(z)$ depending on the position of the galaxy within or outside the eBOSS footprint.\footnote{\texttt{REVOLVER} also uses an estimate of the total volume of the survey in converting normalised Voronoi volumes to true units for determining void sizes; this step is also corrected to account for the position-dependent depth in calculating the volume of the composite survey.}

After applying these changes, the operation of the void-finder results in a catalogue of on average $4700$ voids across both galactic caps (with the exact number identified changing by up to $\pm100$, or $\sim2\%$, depending on the value of $\beta$ used for reconstruction, in the prior range $[0.15,0.55]$). 
The redshift distribution of the resultant voids is shown in Fig.~\ref{fig:nz_sample}, and their size distribution in Fig.~\ref{fig:void_sizes}. 
To these catalogues we apply a size cut, selecting voids with effective radius greater than the median value for the sample for the final analysis. 
This cut corresponds to $R_v>49$ \hmpc{}, and is equivalent to the cut previously applied in the BOSS void analysis \citep{Nadathur:2019c}.
The sample selected for correlation measurements therefore contains on average 2350 voids.
This cut to include only large voids that dominate the dynamics of their local environment is made to ensure the validity of the assumptions made in the modelling, described in Section~\ref{sec:methods} below. 

\subsection{Correlation estimator}
\label{sec:estimator}
The primary observable studied in this work is the void-galaxy correlation function, measured as a function of the observed redshift-space separation between void and galaxy positions, $s$, and the cosine of the angle between the void-galaxy pair and the line of sight $\mu$. 
We estimate this correlation using the Landy-Szalay estimator \citep{Landy:1993}:
\begin{equation}
    \label{eq:LSestimator}
    \xi^s(s,\mu) = \frac{D_1D_2-D_1R_2-D_2R_1+R_1R_2}{R_1R_2}\,
\end{equation}
where each term $XY$ refers to the number of pairs for the given populations in the $(s,\mu)$ separation bin, normalized by the effective total number of such pairs (henceforth we drop the subscript $_\mathrm{vg}$ from $\xi$ as the context is clear). 
Here $D_1$ refers to the void centre positions, $D_2$ to the galaxies, and $R_1$ and $R_2$ to the corresponding sets of random points for the void and galaxy catalogues. 
The galaxy random catalogue $R_2$ is constructed to provide an unclustered set of points, matching the angular and redshift distribution of the observed eBOSS+CMASS galaxies and including appropriate weights to describe observational systematics \citep{Ross20a}, and exceeding the number of galaxies by a factor of 50 (20 for the \ezmocks). 
To construct the appropriate random catalogue $R_1$ matching the void distribution, we first run the void-finding on each of the 1000 \ezmocks{} and use these 1000 catalogues to determine the angular and redshift selection functions of voids. 
We then jointly draw random positions from these distributions to obtain a random catalogue containing $50\times$ as many objects as the DR16 voids. 
To account for potential $\beta$-dependence introduced by reconstruction, we repeat this step to obtain a new void random catalogue for each value of $\beta$ used in the reconstruction.\footnote{However, the $\beta$-dependence was observed to be very small, so this step might have been overkill.}

Using Eq.~\ref{eq:LSestimator} we measure $\xi^s(s,\mu)$ in 80 angular bins $0\leq\mu\leq1$ and 25 equal radial bins $0<s<120\;h^{-1}$Mpc. 
In applying Eq. \ref{eq:LSestimator} we weight the galaxies and galaxy randoms by their associated weights $w_\mathrm{sys,tot}=w_\mathrm{noz}\cdot w_\mathrm{cp}\cdot w_\mathrm{sys}$. 
As systematic effects in the observed galaxy density have already been corrected in the void-finding procedure (Section~\ref{sec:voids}), all voids and void randoms are given equal unit weights by default. 
No significant differences were seen between the NGC and SGC samples, so we always present results for the combined correlation obtained by adding pair counts in the two caps.

In order to estimate the redshift-space correlation $\xi^s$ we use the galaxy positions in redshift space; for the \ezmocks{} we also estimate the real-space correlation $\xi^r$ using Eq.~\ref{eq:LSestimator} by simply replacing these with the real-space galaxy positions determined from the reconstruction step. 
In both cases, the void catalogues used are always those determined from the post-reconstruction, RSD-removed galaxy field. Therefore both $\xi^s$ and $\xi^r$ always have an implicit dependence on $\beta$.

We decompose the measured correlation function into its Legendre multipoles $\xi_\ell(s)$ as
\begin{equation}
    \label{eq:Legendre}
    \xi_\ell(s)\equiv\frac{2\ell+1}{2}\int_{-1}^{1}L_\ell(\mu)\xi(s,\mu)\,\mathrm{d}\mu\,,
\end{equation}
where $L_\ell(\mu)$ is the Legendre polynomial of order $\ell$, and we are interested in only the monopole and quadrupole, $\ell=0,2$. 
Fig.~\ref{fig:data_vs_mocks} shows the redshift-space monopole $\xi^s_0$ and quadrupole $\xi^s_2$ in the eBOSS+CMASS data compared to those from the mean of the \ezmock, \nseries{} and \patchy{} mocks.

\subsection{Covariance matrix}
\label{sec:covmat}
To estimate the uncertainties for our measurement on the LRG data, we use the \ezmocks, constructing the covariance matrix for the individual bin measurements as
\begin{equation}
    \label{eq:covmat}
    \mathbf{C} = \frac{1}{999}\sum_{k=1}^{1000}\left(\boldsymbol{\xi}^k-\overline{\boldsymbol{\xi}^k}\right)\left(\boldsymbol{\xi}^k-\overline{\boldsymbol{\xi}^k}\right)\,,
\end{equation}
where $\boldsymbol{\xi}=\left(\xi^s_0, \xi^s_2\right)$ denotes the data vector composed of the binned measurement of monopole and quadrupole moments, $k$ is the index identifying the individual \ezmock{} realization, and $\overline{\boldsymbol{\xi}^k}$ is the mean data vector over the 1000 mocks. 
The same method applied to the NGC samples of the \patchy{} mocks is used to generate the covariance matrix for the \nseries.

As previously noted, the covariance matrix inherits a dependence on $\beta$ from the data vector as a result of the $\beta$-dependent reconstruction and RSD removal process. This is accounted for as described in Section \ref{sec:methods} below.

\subsection{Effective redshift}
\label{sec:effz}
We define the effective redshift of our void-galaxy measurement using the weighted sum
\begin{equation}
    \label{eq:effz}
    z_\mathrm{eff} = \frac{\sum_{ij}\left(\frac{Z_i + z_j}{2}\right)w_j}{\sum_{ij} w_j}\,,
\end{equation}
where $Z_i$ and $z_j$ are the individual void and galaxy redshifts, respectively, $w_j=w_\mathrm{sys,tot}$ is the associated galaxy systematic weight, and the sum extends over all void-galaxy pairs included in the computation of the correlation. Here the void redshift $Z_i$ is calculated as the cosmological redshift that would be expected for a hypothetical object located at the void centre position.

For the DR16 LRG data, after void selection cuts we find the effective redshift is $z_\mathrm{eff}=0.690$, with negligible dependence on $\beta$ and no difference between the NGC and SGC samples. In Section \ref{sec:consensus} below, we combine our void-galaxy measurements with those from the consensus BAO+RSD galaxy clustering results from the same sample \citep{LRG_corr,gil-marin20a}, which have an effective redshift $z_\mathrm{eff}=0.698$. 
As these two values are very similar to each other, in doing so we will simply ignore the difference and treat the consensus results as applying at the effective redshift $z_\mathrm{eff}=0.70$. 

For the AP measurements, the effective redshift is used to interpret measurements taken across the survey scaled to a common basis (because comoving units are used), as a single measurement at a particular redshift. The results are therefore largely insensitive to the exact value of the effective redshift.  
RSD measurements in principle depend more strongly on the effective redshift as the amplitude of clustering measured over the full survey is not adjusted to a common baseline before being averaged (by calculating the correlation function summed across the sample) but is compared to a model calculated at the effective redshift. However, the expected change in growth rate $f\sigma_8$ over the range of effective redshifts considered ($0.69<z_\mathrm{eff}<0.70$) is negligible compared with the measurement error.

%%%%%% MODEL FITTING %%%%%%
\section{Model fitting}
\label{sec:methods}

\subsection{Model}
\label{sec:model}
We use the linear dispersion model of \citet{Nadathur:2019a} to describe the redshift-space void-galaxy correlation function $\xi^s(\mathbf{s})$. According to this model, in the true cosmology, $\xi^s(\mathbf{s})$ is related to the (spherically symmetric) real-space correlation $\xi^r(r)$ by
\begin{equation}
    \label{eq:full_model}
    1 + \xi^s(s, \mu) = \int\left(1 + \xi^r(\tilde{r})\right)\left[1 + \frac{\tilde{v}_r}{\tilde{r}aH} + \frac{\tilde{r}\tilde{v}_r^\prime - \tilde{v}_r}{\tilde{r}aH}\mu^2\right]^{-1}P(v_{||})\,\rmn{d}v_{||},
\end{equation}
where the $^\prime$ denotes the derivative with respect to $r$,  
\begin{equation}
    \label{eq:r_tilde}
    \tilde{r} = \sqrt{r_\perp^2 + \left(r_{||}-\tilde{v}_r/aH\right)^2}\,,
\end{equation}
for real-space void-galaxy separation distances $r_\perp$ and $r_\parallel$ perpendicular and parallel to the line of sight direction respectively, 
\begin{equation}
    \label{eq:v_tilde}
    \tilde{v}_r = v_r - v_{||}\mu
\end{equation}
is the radial component of the galaxy peculiar velocity relative to the void centre, 
\begin{equation}
    \label{eq:pdf}
    P(v_\parallel) = \frac{1}{\sqrt{2\pi}\sigma_{v_\parallel}(r)} \exp\left(-\frac{v_\parallel^2}{2\sigma_{v_\parallel}^2(r)} \right)
\end{equation}
is a Gaussian pdf for the random line-of-sight velocity component $v_\parallel$ described by a position-dependent dispersion function $\sigma_{v_\parallel}(r)$, and $v_r(r)$ is the coherent radially-directed galaxy outflow velocity relative to the void centre. 
In this model the mapping between real-space and redshift-space coordinates is described by
\begin{equation}
    \label{eq:coordinates}
    s_\perp = r_\perp\;;\;s_\parallel = r_\parallel + \frac{\tilde{v}_r(r)}{aH}\,,
\end{equation}
where $a$ is the scale factor and $H$ the Hubble rate at the redshift of the void.

We will assume only linear perturbation theory results here, as this has been shown to provide an excellent description of the void-galaxy correlation on all separation scales in this model \citep{Nadathur:2019a,Nadathur:2019c}. 
Under this assumption, and further assuming that the galaxy peculiar velocities are determined by the void alone, 
\begin{equation}
    \label{eq:vr}
    v_r(r) = -\frac{1}{3}faHr\Delta(r)\,,
\end{equation}
where $f$ is the linear growth rate and $\Delta(r)$ is the average mass density contrast within radius $r$ of the void centre, defined in terms of the void matter density profile $\delta(r)$ as
\begin{equation}
    \label{eq:Delta}
    \Delta(r) = \frac{3}{r^3}\int_0^r\delta(y)y^2\,\rmn{d}y\,.
\end{equation}

In the limit of zero velocity dispersion, $\sigma_{v_\parallel}\rightarrow0$, Eq.~\ref{eq:full_model} reduces to the expression 
\begin{equation}
    \label{eq:Kaiser limit}
    1 + \xi^s(s, \mu) = \left(1 + \xi^r(r)\right)\left[1 + \frac{v_r}{raH} + \frac{rv_r^\prime - v_r}{raH}\mu^2\right]^{-1}
\end{equation}
derived by \citet{Cai:2016a}. 
This is the equivalent of the Kaiser RSD model for galaxy clustering \citep{Kaiser:1987}, with the term in square brackets representing the Jacobian of the coordinate transformation under the mapping $\mathbf{r}\rightarrow\mathbf{s}$. 
An approximation to Eq.~\ref{eq:Kaiser limit}, obtained by substituting for $v_r$ using Eq.~\ref{eq:vr}, expanding the square brackets and dropping terms of order $\xi^r\Delta$ and $\Delta^2$ or higher, has been used in some other works \citep[e.g.,][] {Cai:2016a,Hamaus:2017a,Achitouv:2019,Hawken:2020,Aubert20a}. 
Under this approximation, Eq.~\ref{eq:Kaiser limit} reduces to
\begin{equation}
    \label{eq:linear_approx}
    \xi^s(r,\mu) = \xi^r(r) + \frac{f\Delta(r)}{3} + f\mu^2\left(\delta(r)-\Delta(r)\right)\,,
\end{equation}
which is referred to by these authors as a ``linear model" (note that here the approximation $s\simeq r$ is also used as in the cited works). 
However, the validity of the approximation used to derive  Eq.~\ref{eq:linear_approx} has been questioned by \citet{Nadathur:2019a}, as terms of order $\xi^r\Delta$ are generally not negligible compared to $\Delta$. 
Our baseline model avoids this approximation by directly evaluating the terms in square brackets in Eq.~\ref{eq:full_model} (or Eq.~\ref{eq:Kaiser limit}) exactly, without the need to truncate a series expansion at any order. We compare the performance of various models in Appendix~\ref{sec:appendixA}.

\begin{figure}
    \centering
    \includegraphics[width=0.8\columnwidth]{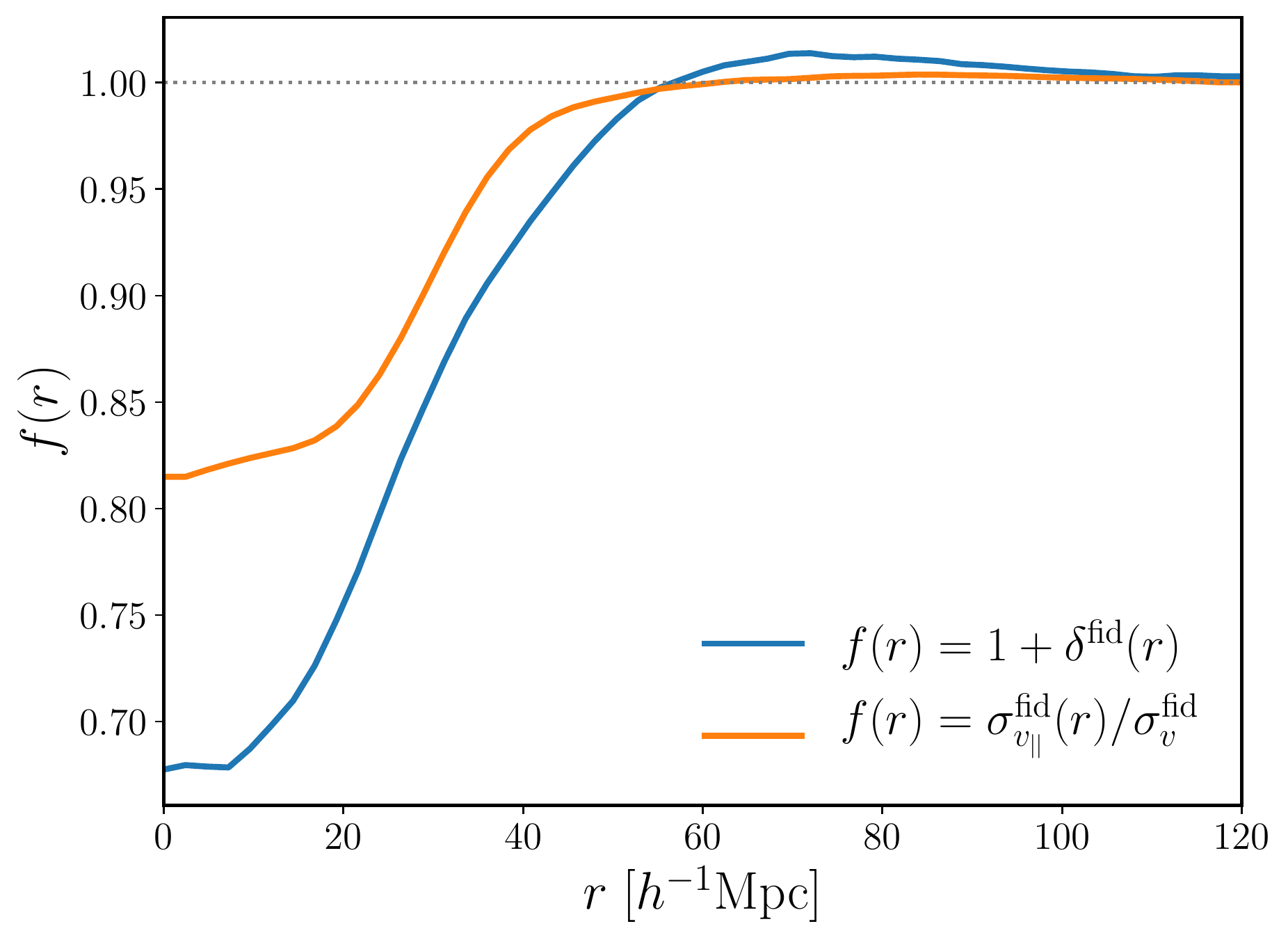}
    \caption{The void matter density profile $\delta(r)$ and galaxy velocity dispersion profile $\sigma_{v_\parallel}(r)$ measured from the eBOSS LRG-like mock galaxy and void catalogues in the \bigmd{} simulation at $z=0.70$. 
    Interpolations to these functions are used to calibrate the model of the void-galaxy correlation before application to the data as described in Section~\ref{sec:model}.
    } 
    \label{fig:calibration}
\end{figure}

In order to obtain model predictions from Eq.~\ref{eq:full_model}, the input functions $\xi^r(r)$, $\delta(r)$ and $\sigma_{v_\parallel}(r)$, need to be specified. 
To do this we adopt the procedure followed by \citet{Nadathur:2019c}. 
We use the dark matter particle output and the mock galaxy velocities in the \bigmd{} $N$-body simulation to measure the $\delta(r)$ and $\sigma_{v_\parallel}(r)$ profiles for voids in the \bigmd{} mock at redshift 0.70, denoted as $\delta^\mathrm{fid}(r)$ and $\sigma_{v_\parallel}^\mathrm{fid}(r)$ respectively. 
The procedure for estimating these functions follows that outlined by \citet{Nadathur:2019a}, and the calibrated functions are shown in Fig.~\ref{fig:calibration} (with $\sigma_{v_\parallel}^\mathrm{fid}(r)$ shown normalised in units of its asymptotic amplitude far from the void centre, $\sigma_v^\mathrm{fid}\equiv\sigma_{v_\parallel}^\mathrm{fid}(r\rightarrow\infty)$). 
When calculating model predictions from Eq.~\ref{eq:full_model} we then substitute
\begin{equation}
    \label{eq:delta}
    \delta(r) = \frac{\sigma_8(z)}{\sigma_8^\mathrm{MD}(0.70)}\delta^\mathrm{fid}(r)\,,
\end{equation}
where $\sigma_8^\mathrm{MD}(0.70)=0.579$ is the $\sigma_8$ value for the \bigmd{} simulation at redshift 0.70, and  
\begin{equation}
    \label{eq:sigmav}
    \sigma_{v_\parallel}(r) = \frac{\sigma_v}{\sigma_v^\mathrm{fid}} \sigma_{v_\parallel}^\mathrm{fid}(r)\,,
\end{equation}
with $\sigma_v$ taken to be a free parameter. 
Note that the model calibration is performed using the \bigmd{} mock, which is entirely independent of the \ezmocks, \nseries{} mocks and of course the DR16 data that are fit using the resultant model. 
This calibration procedure does mean that we are essentially performing a template fit to the RSD seen in the data, using a template derived from a fiducial \lcdm{} cosmology. 
\citet{Nadathur:2019c} investigated the changes in the derived template using a range of different simulations, confirming the assumed linear scaling with $\sigma_8$ in Eq.~\ref{eq:delta}.
While our approach is similar in spirit to the template fitting used in state-of-the-art galaxy clustering analyses \citep[e.g.,][]{Beutler:2017,gil-marin20a,LRG_corr}, further work is desirable to investigate potential limitations to its use for models that are far from the fiducial cosmology.

Finally, the real-space correlation monopole $\xi^r(r)$ can be measured directly from the RSD-removed galaxy field (with an implicit $\beta$-dependence, of course). 
However, as this measurement on the DR16 data is noisy and the noise is correlated with that in the measurement of $\xi^s$, this can lead to unnaturally small $\chi^2$ values. We therefore use the mean $\xi^r(r)$ measured over the 1000 \ezmocks{} instead. We have checked that this choice does not affect the posterior parameter estimates. 

The model of Eq.~\ref{eq:full_model} is extended to account for possible differences between the true cosmology and the fiducial model used to convert redshifts to distances by introducing the Alcock-Paczynski (AP) $\alpha$ parameters
\begin{equation}
    \label{eq:alphas}
    \aperp\equiv\frac{D_M(z)}{D_M^\mathrm{fid}(z)}\;;\;\apar\equiv\frac{D_H(z)}{D_H^\mathrm{fid}(z)}\,,
\end{equation}
where $D_M(z)$ is the transverse comoving distance and $D_H(z) = c/H(z)$ is the Hubble distance at redshift $z$. 
With this notation, 
\begin{equation}
    \label{eq:model_with_alphas}
    \xi^s(s_\perp, s_\parallel) = \xi^{s,\mathrm{fid}}\left(\aperp s_\perp^\mathrm{fid}, \apar s_\parallel^\mathrm{fid}\right)\,.
\end{equation}
In order to avoid accidentally introducing a preferred fundamental void size scale to the model, whenever $\apar$ and $\aperp$ differ from unity we rescale the input functions $\xi^r(r)$, $\delta(r)$ and $\sigma_{v_\parallel}(r)$ as described by \citet{Nadathur:2019c}. 
This rescaling is equivalent to changing the distance argument in these functions by $r\rightarrow\aperp^{2/3}\apar^{1/3}r$, i.e. dilating the apparent void size. 
It therefore means that the final model is sensitive only to the ratio $\aperp/\apar$ of the AP $\alpha$ parameters. 
Unlike the case for BAO analyses, we have defined our AP $\alpha$ parameters here without reference to the sound horizon scale $r_d$, although as we are only sensitive to their ratio this makes no practical difference.

%%%%%% FIGURE %%%%%%

\subsection{Parameter inference}
\label{sec:inference}
The model described in Section \ref{sec:model} above depends on four free parameters: $f\sigma_8$, $\aperp/\apar$, $\beta$ and $\sigma_v$. The definition of the density profile calibrated from the \bigmd{} mock, Eq.~\ref{eq:delta}, has introduced a degeneracy between $f$ and $\sigma_8$ so that, as for standard RSD analyses, the model is sensitive only to the combination $f\sigma_8$. 
The dependence on $\beta=f/b$ arises because of the implicit dependence of $\xi^r(r)$ and the data vector on $\beta$ via the RSD removal before the void-finding. 
Following the procedure of \citet{Nadathur:2019c}, in order to avoid the growth rate of cosmological interest appearing in two separate parameters in different combinations, we simply change parameter basis to $\left(f\sigma_8, \aperp/\apar, \sigma_v, b\sigma_8\right)$ when exploring the posterior. 
We then treat $\sigma_v$ and $b\sigma_8$ as nuisance parameters that are not of interest for cosmology, and always marginalize over them when reporting results on $f\sigma_8$ and $\aperp/\apar$.

In order to explore the posterior over this parameter space we make use of the public code \texttt{Victor}\footnote{VoId-galaxy CorrelaTion cOsmology fitteR, \url{https://github.com/seshnadathur/victor}. The name of this code was generated using \texttt{acronym} \citep{Cook:2019}.}, which is a general-purpose void-galaxy correlation code designed to implement several alternative models for the multipoles $\xi^s_0$ and $\xi^s_2$ and perform posterior fits. 
At each point in the parameter space, we calculate
\begin{equation}
    \label{eq:chi2}
    \chi^2 = \left(\boldsymbol{\xi}^{s,\mathrm{th}}-\boldsymbol{\xi}^s\right) \mathbf{C}^{-1} \left(\boldsymbol{\xi}^{s,\mathrm{th}}-\boldsymbol{\xi}^s\right)\,,
\end{equation}
where $\boldsymbol{\xi}^{s,\mathrm{th}}$ is the theory data vector calculated from the model of Eq. \ref{eq:full_model} and $\mathbf{C}$ is the covariance matrix estimated from Eq. \ref{eq:covmat}. 
To correctly propagate the uncertainty in the covariance matrix estimation through to the parameter inferences we use the full likelihood described by \citet{Sellentin:2016}, 
\begin{equation}
    \label{eq:likelihood}
    \ln\mathcal{L}=-\frac{N}{2}\ln\left(1+\frac{\chi^2}{N-1}\right) - \frac{\det{\mathbf{C}}}{2}\,,
\end{equation}
where $N=1000$ is the number of \ezmocks{} used to estimate $\mathbf{C}$, and the $\det{\mathbf{C}}/2$ normalization term explicitly accounts for the fact that the covariance matrix varies with $\beta$. 
We impose uninformative flat priors $f\sigma_8\in[0.05, 1.5]$, $b\sigma_8\in[0.1, 2]$, $\beta\in[0.15, 0.55]$, $\aperp/\apar\in[0.8, 1.2]$ and $\sigma_{v_{||}}\in[100, 600]$, and explore the posterior distribution using the affine-invariant ensemble MCMC sampler \texttt{emcee} \citep{Foreman-Mackey:2013} implemented in \texttt{Victor}. 
In order to make the MCMC exploration feasible in finite time we precompute the data vector $\boldsymbol{\xi}^s$, the covariance matrix $\mathbf{C}$ and the input function $\xi^r$ on a grid of $\beta$ chosen to efficiently explore a wide prior range $\beta\in\left[0.15, 0.55\right]$ around the expected value $\beta^\mathrm{exp}=0.353$ for a \lcdm{} cosmology with bias $b=2.3$ as for the \ezmocks. 
When running the chains we interpolate on this grid to obtain these quantities at any intermediate values of $\beta$. 
We run four independent \texttt{emcee} chains, each with 100 walkers, until the length of each chain exceeds 2000 times the maximum autocorrelation length in any parameter. 
The Gelman-Rubin convergence criterion \citep{Gelman:1992} for the final set of four chains is $R-1=0.002$.

%%%%%% FIGURE %%%%%%
\begin{figure*}
    \centering
    \includegraphics[width=0.95\linewidth]{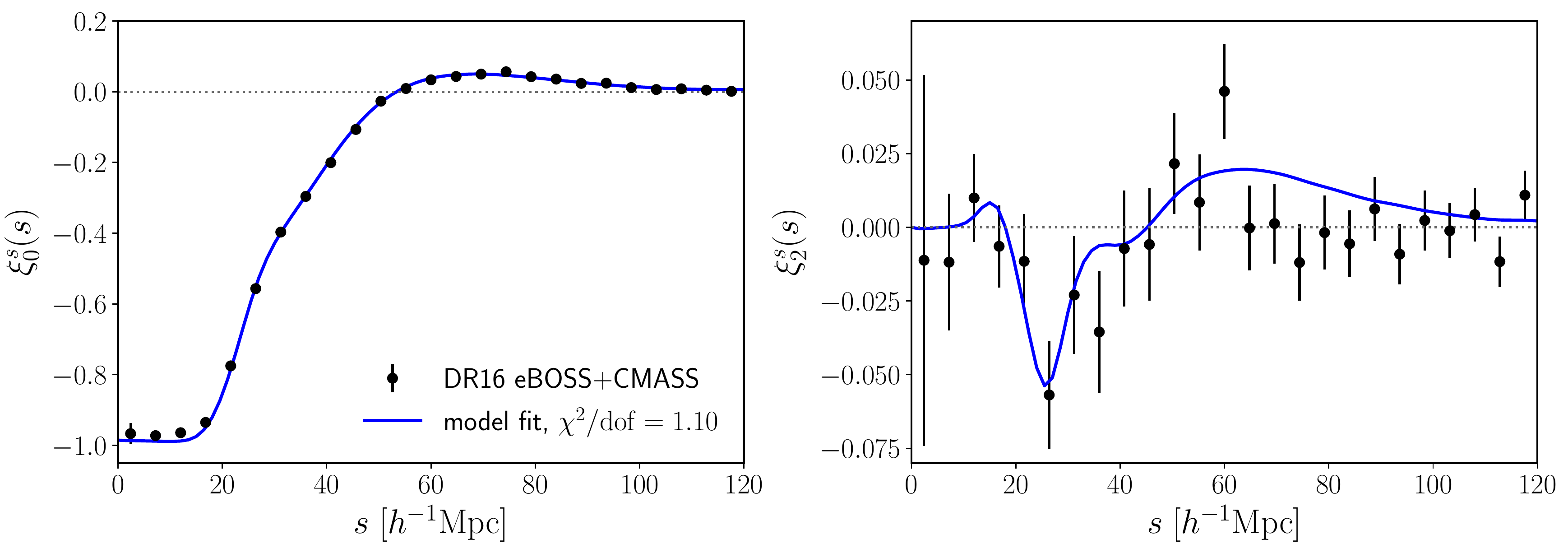}
    \caption{Monopole (left) and quadrupole (right) moments of the void-galaxy correlation measured in the DR16 eBOSS+CMASS joint LRG sample and the best-fit theory model of Eq.~\ref{eq:full_model} describing these data, corresponding to parameter values $f\sigma_8=0.355$ and $\aperp/\apar=1.011$ at the maximum likelihood point. 
    Error bars on the data points are the $1\sigma$ uncertainties derived from the diagonal entries of the covariance matrix calculated from the 1000 \ezmocks.
    } 
    \label{fig:dr16multipoles}
\end{figure*}

%%%%%% FIGURE %%%%%%
\begin{figure}
    \centering
    \includegraphics[width=\columnwidth]{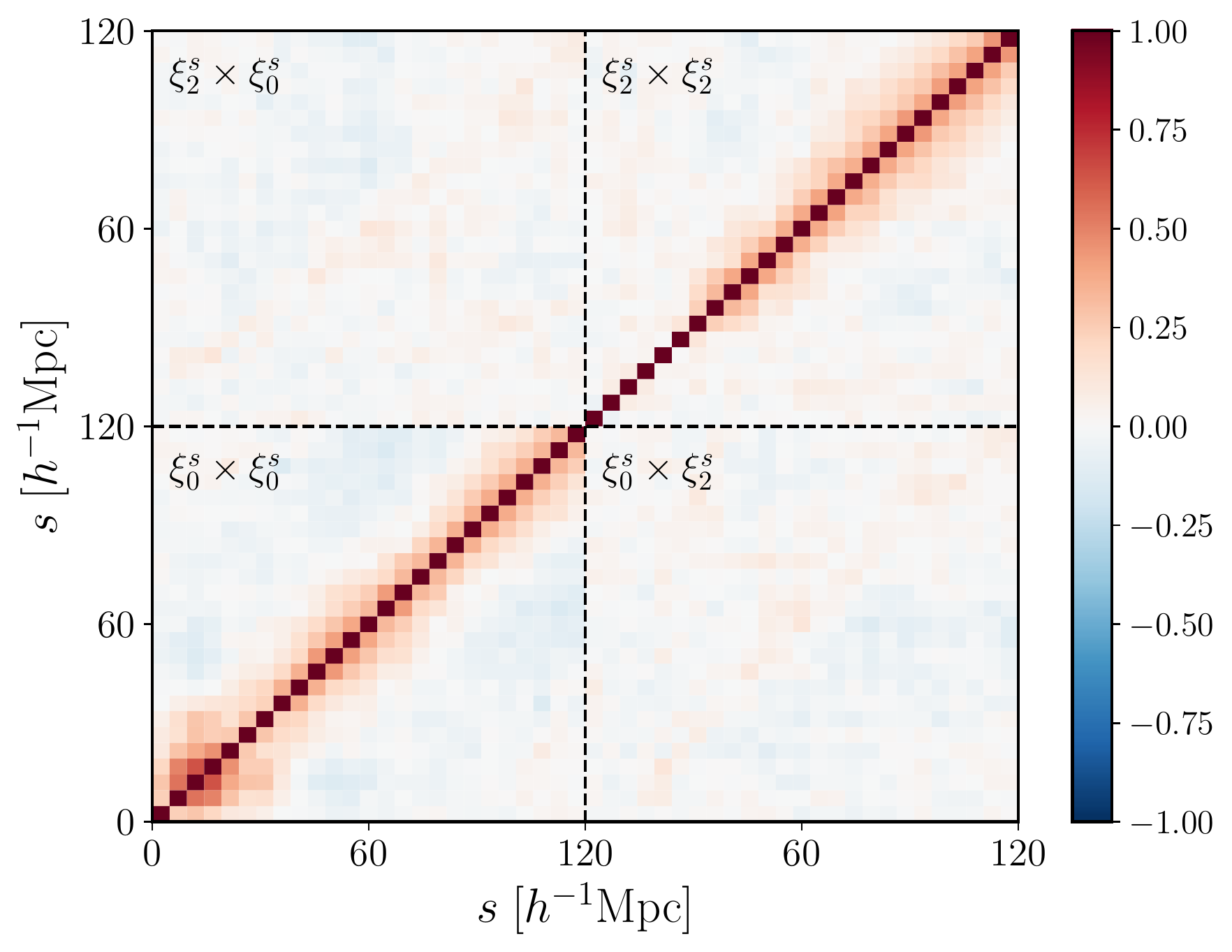}
    \caption{The normalised correlation structure of the covariance matrix for the redshift-space correlation multipole measurement derived from the 1000 \ezmocks. 
    The covariance matrix has a very small dependence on $\beta$, which is accounted for in the model-fitting; this plot shows the covariance at the value $\beta=0.35$ for illustration.
    } 
    \label{fig:corrmat}
\end{figure}

%%%%%% FIGURE %%%%%%
\begin{figure}
    \centering
    \includegraphics[width=0.95\columnwidth]{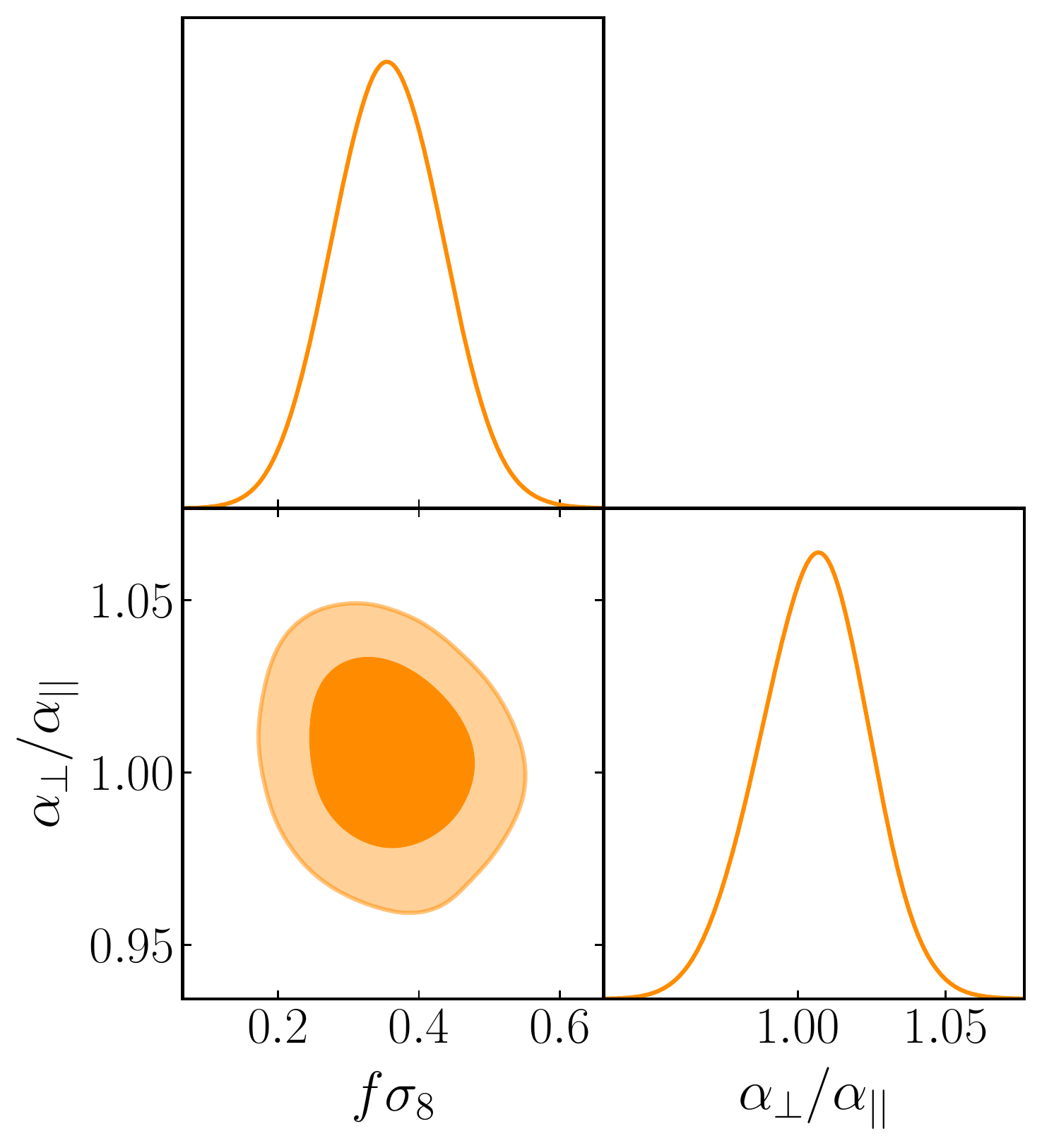}
    \caption{Marginalised posterior constraints on the cosmological parameters $f\sigma_8$ and $\aperp/\apar$ from the fit to the DR16 data. 
    The shaded contours show the $68\%$ and $95\%$ confidence limit regions. 
    This plot includes only statistical errors. 
    The contribution of systematic errors in each parameter is assessed in Section~\ref{sec:systematics} and added in quadrature in the final results reported in Eqs.~\ref{eq:void_fs8_result} and \ref{eq:void_DmDh_result}. 
    } 
    \label{fig:fs8_FAP}
\end{figure}

%%%%%% RESULTS %%%%%%
\section{Results}
\label{sec:results}
In Fig.~\ref{fig:dr16multipoles} we show the comparison between the DR16 eBOSS+CMASS void-galaxy correlation data and the best-fit model of  Eq.~\ref{eq:full_model}. 
The model provides an excellent fit to the data, with a $\chi^2$ of $50.6$ for $(50-4)$ degrees of freedom. 
The correlation structure of the covariance matrix for the data vector is shown in Fig. \ref{fig:corrmat}, and shows a generally diagonal structure with a small degree of correlation between neighbouring data bins.

Fig.~\ref{fig:fs8_FAP} shows the resultant marginalised constraints obtained on the model parameters $f\sigma_8$ and $\aperp/\apar$ from the void data. 
The marginalised 1D constraints on each parameter are $f\sigma_8 = 0.356\pm 0.077$ and $\alpha_\perp/\alpha_{||} = 1.005\pm 0.018$ (statistical errors only), with a weak negative correlation between them (correlation coefficient of $\rho=-0.154$). 

The systematic error contributions to these measurements are studied below in Section~\ref{sec:systematics}, and are determined to be $\sigma_{f\sigma_8}^\mathrm{syst}=0.016$ and $\sigma_{\aperp/\apar}^\mathrm{syst}=0.009$. 
Using these and the values of $D_M^\mathrm{fid}$ and $D_H^\mathrm{fid}$ for our fiducial cosmological model, we can summarize the final cosmological results of our void-galaxy measurement as
\begin{equation}
    \label{eq:void_fs8_result}
    f\sigma_8 = 0.356 \pm 0.079
\end{equation}
and
\begin{equation}
    \label{eq:void_DmDh_result}
    \frac{D_M}{D_H} = 0.868 \pm 0.017
\end{equation}
at effective redshift $z_\mathrm{eff}=0.69$. 
The constraint on the distance ratio can be rephrased in terms of one on the Alcock-Paczynski distortion parameter
\begin{equation}
    \label{eq:epsilon}
    \epsilon \equiv\left(\frac{\apar}{\aperp}\right)^{1/3}-1 = -0.0017 \pm 0.0067\,,
\end{equation}
consistent with \lcdm, and a significantly higher precision measurement than that obtained from BAO.

%%%%%% SYSTEMATICS %%%%%%
\section{Systematics tests}
\label{sec:systematics}
In this Section we check the robustness of our pipeline and quantify the systematic error budget of our measurement. 
In particular we wish to determine the contribution to the total error budget of
\begin{itemize}
    \item \emph{Modelling systematics}, or those errors associated with inaccuracies or limitations of the theoretical model of Eq.~\ref{eq:full_model}, or with its applicability to the data in question. 
    To determine this we run all the steps of our pipeline on both the \ezmock{} and \nseries{} mock catalogues, analysed in the $\Omega_m=0.31$ and $\Omega_m=0.286$ cosmological models respectively, which we take as the true cosmologies of these mocks (the difference between $\Omega_m=0.31$ and $\Omega_m^\mathrm{\textsc{EZmocks}}=0.307$ being very small).
    \item \emph{Fiducial cosmology systematics}, or those errors associated with performing the analysis in a cosmological model that differs from the true cosmology of the data. 
    To test this, we use the \nseries{} mocks and repeat the whole analysis pipeline in two cosmological models that differ from the true cosmology of the mocks, with $\Omega_m=0.310$ and $\Omega_m=0.350$. 
    The second of these models is specifically chosen to be far from the true cosmology of the \nseries{} mocks, $\Omega_m^\mathrm{\nseries}=0.286$, and to be strongly disfavoured by state-of-the-art CMB analyses \citep{Planck:2018params}. 
    It could therefore be regarded as a somewhat extreme case.
\end{itemize}

Note that the \ezmocks{} are approximate and, unlike the full $N$-body \nseries{} mocks, they are not expected to reproduce the correct RSD to percent-level accuracy. 
On the other hand, the \nseries{} mocks represent the CMASS galaxy sample only, which is more homogeneous than the composite eBOSS+CMASS sample and does not require the corrections to the void-finding algorithm described in Section~\ref{sec:voids}. 
Therefore we perform tests for modelling systematics on both sets of mocks, and conservatively choose the larger of the observed offsets in the two cases to represent the systematic error on each parameter.  

\begin{table*}
  \centering
  \caption{Performance of the model of Eq.~\ref{eq:full_model} when analysing the mocks in their own cosmology (for \nseries) or very close to their own cosmology (\ezmocks). 
  Here $(f\sigma_8)^\mathrm{exp}$ and $(\aperp/\apar)^\mathrm{exp}$ are the expected values of the parameters for these mocks. 
  We show the differences between the mean value obtained from the mocks and these expected values as $\Delta(f\sigma_8)$ and $\Delta(\aperp/\apar)$ respectively. 
  The $2\sigma$ uncertainties on these differences are determined as twice the mean of the 1D marginalized parameter uncertainties in the individual mocks multiplied by $1/\sqrt{N_\mathrm{mocks}}$.
  }
  \begin{tabular}{lcccccc}
    \hline
    \hline
    Mock & $N_\mathrm{mocks}$ & ref. cosmology & $(f\sigma_8)^\mathrm{exp}$ & $(\aperp/\apar)^\mathrm{exp}$ & $\Delta(f\sigma_8)\pm2\sigma$ & $\Delta(\aperp/\apar)\pm2\sigma$ \\ 
    \hline
    \ezmocks & 1000 & $\Omega_m=0.310$ & 0.4687 & 0.99871 & $-0.0133\pm0.0048$ & $0.00076\pm0.00097$ \\
    \nseries & 84 & $\Omega_m=0.286$ & 0.4703 & 1.00000 & $-0.0141\pm0.0144$ & $-0.00417\pm0.00390$ \\
    \hline
  \end{tabular}
  \label{table:model_err}
\end{table*}

%%%%%% FIGURE %%%%%%
\begin{figure}
    \centering
    \includegraphics[width=\columnwidth]{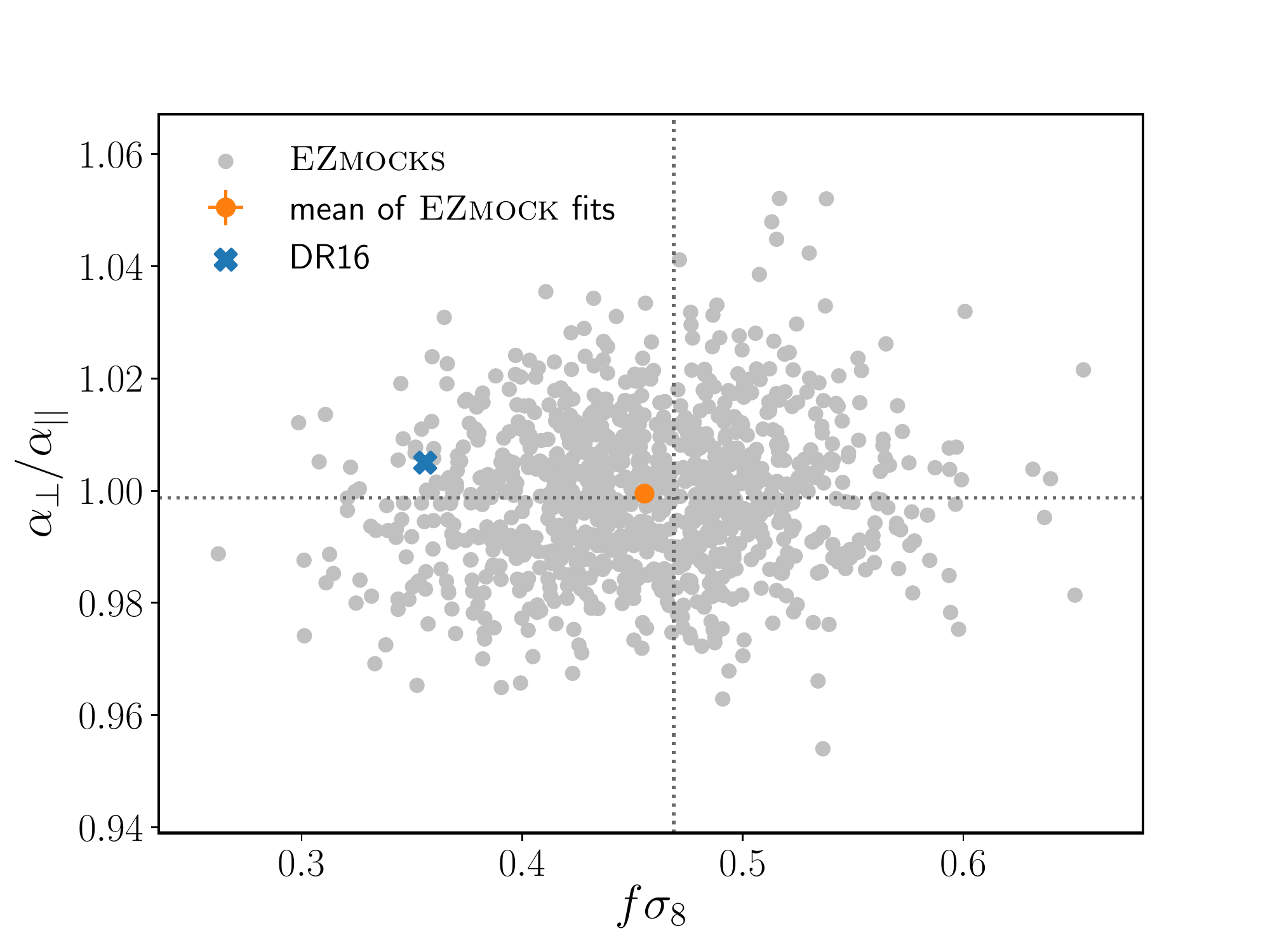}
    \caption{Performance of the void-galaxy model applied to the \ezmocks{} at $\Omega_m=0.310$. 
    The grey points show the recovered mean values of the parameters $f\sigma_8$ and $\aperp/\apar$ in each of the 1000 \ezmock{} realisations. 
    The orange circle shows the mean of these values and the error bars in each direction represent the error in the mean (which is $\sqrt{1/1000}$ times the mean error in an individual mock). 
    The dashed lines indicate the true expected values $(f\sigma_8)^\rmn{exp}=0.4687$ and $(\aperp/\apar)^\rmn{exp}=0.9987$ respectively. 
    The blue cross shows the result of the measurement on the actual DR16 eBOSS+CMASS sample.
    } 
    \label{fig:syst_fs8_FAP}
\end{figure}

%%%%%% FIGURE %%%%%%
\begin{figure}
    \centering
    \includegraphics[width=\columnwidth]{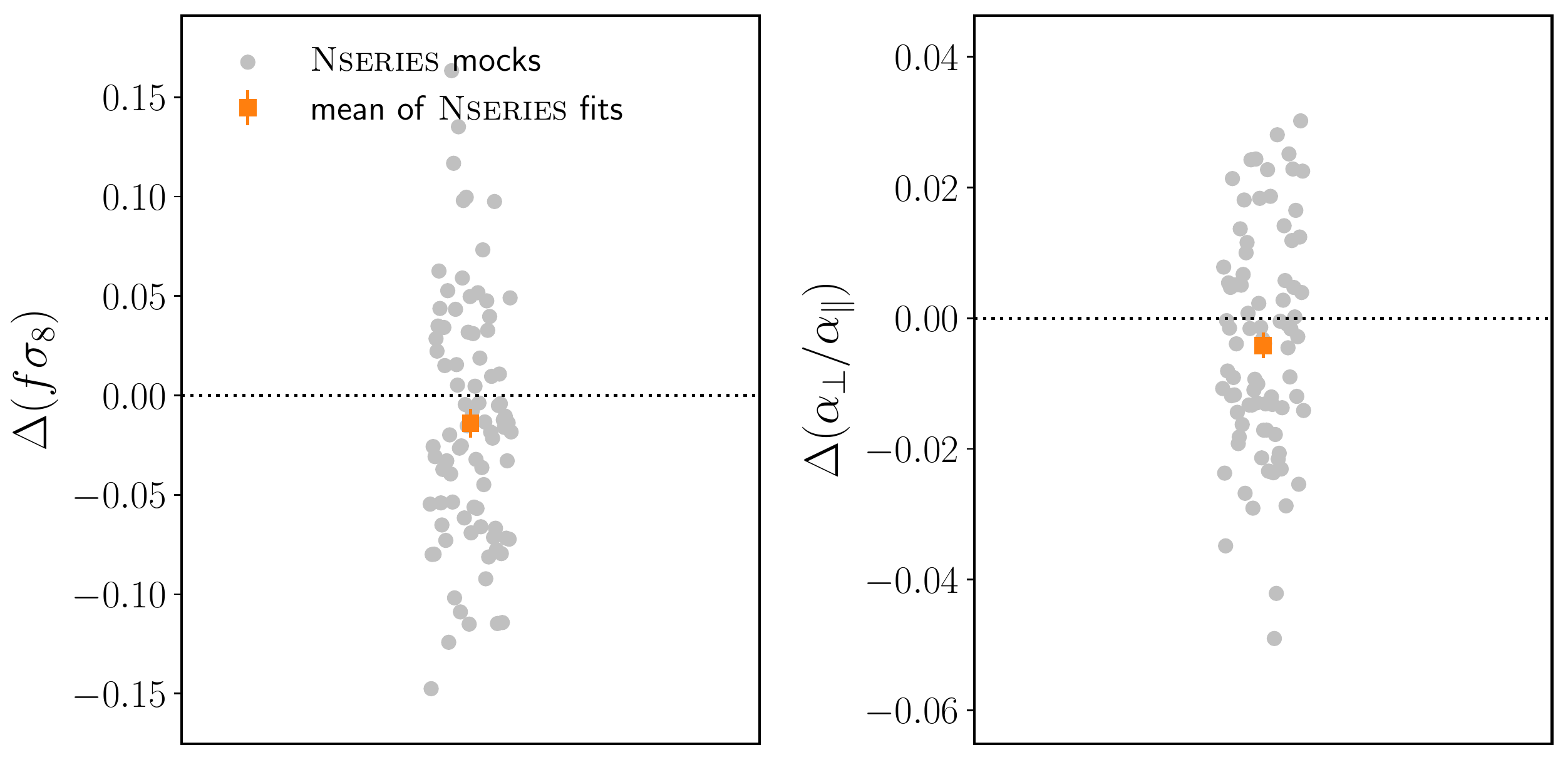}
    \caption{Performance of the void-galaxy model applied to the \nseries{} mocks at $\Omega_m=0.286$. 
    The vertical axes show the difference between the recovered and true values of $f\sigma_8$ (left panel) and $\aperp/\apar$ (right panel). 
    The grey points represent the recovered parameter values for each of the 84 mocks. 
    The orange squares represent the mean of these measurements, with error bars representing the error in the mean, calculated as $\sqrt{1/84}$ times the mean error in a single mock. 
    } 
    \label{fig:syst_nseries}
\end{figure}

\subsection{Test of modelling errors in the true cosmology}
\label{sec:model_syst}
Scatter plots of the parameters inferred from applying our model to the individual \ezmock{} and \nseries{} mock catalogues in the true cosmologies are shown in Figs.~\ref{fig:syst_fs8_FAP} and \ref{fig:syst_nseries} for the \ezmocks{} and \nseries{} mocks respectively. 
The results of the tests for modelling systematics are summarized in Table~\ref{table:model_err}. In the \ezmocks{} we see a small but statistically significant offset in the recovered growth rate, with $\Delta(f\sigma_8)=-0.0133\pm0.0048$ ($2\sigma$ uncertainty limit). 
For the \nseries{} the offset is not statistically significant at the $2\sigma$ level, but the statistical resolution of the 84 \nseries{} mocks is limited at a level above that of the offset seen in the \ezmocks. 
As a conservative estimate, we take the $2\sigma$ limit of the resolution of the \nseries{} mocks to represent the modelling systematic contribution to the growth rate uncertainty, $\Delta(f\sigma_8)=0.0144$. 
This corresponds to a $3\%$ effect compared to the expected value for the \nseries.

For the distance ratio $\aperp/\apar$ the systematic offset in the \ezmocks{} is constrained to be less than 0.00097, or $<0.01\%$, completely consistent with zero. 
However, for the lower statistical resolution of the \nseries{} mocks we see marginal evidence of a systematic shift $\Delta(\aperp/\apar)=0.00417\pm0.00390$, or $0.42\%$. 
As before we conservatively take this larger value to be the modelling systematic contribution to $\aperp/\apar$.

Fig.~\ref{fig:syst_fs8_FAP} also shows that the result obtained from the DR16 data is not a significant outlier with respect to the scatter observed in the \ezmocks, with the $f\sigma_8$ value obtained being lower than the mean of the mock values by approximately $1.7\times$ the rms of the scatter, and the value of $\aperp/\apar$ being consistent with that of the mocks.

Although in this work we do not consider the galaxy bias a fundamental cosmological parameter of interest and always marginalise over it in our main analysis, we note in passing that the values of $b\sigma_8$ recovered from this method are $b\sigma_8=1.220\pm0.014$ and $b\sigma_8=1.224^{+0.019}_{-0.023}$ for the \ezmocks{} and \nseries{} mocks, respectively, which are also in line with the expected values for these samples obtained from fitting to the galaxy two-point statistics.

\begin{table}
  \centering
  \caption{Additional contribution to systematic errors from the choice of fiducial cosmology used in analysis of the \nseries{} mocks. 
  Differences are quoted as the additional systematic offset in each parameter compared to the reference values in Table \ref{table:model_err} when the mocks are analysed in their own true cosmology.
  }
  \begin{tabular}{cccc}
    \hline
    \hline
Mock & ref. cosmology & $\Delta(f\sigma_8)^\mathrm{ref}$ & $\Delta(\aperp/\apar)^\mathrm{ref}$ \\ 
\hline
\nseries & $\Omega_m=0.310$ & 0.0075 & 0.00071 \\
\nseries & $\Omega_m=0.350$ & 0.0047 & 0.00810 \\

\hline
  \end{tabular}
  \label{table:ref_cosmo_syst}
\end{table}

\begin{table*}
  \centering
  \caption{Summary of the systematic error budget for the cosmological parameters $f\sigma_8$ and $\aperp/\apar$ determined from the mocks. 
  Modelling contributions to the systematic errors are taken from the worst case offsets for the \nseries{} mocks and \ezmocks{} in Table~\ref{table:model_err}. 
  Cosmology errors are from the worst case offsets in Table~\ref{table:ref_cosmo_syst} when the \nseries{} mocks are analysed using different fiducial cosmologies. 
  Error contributions from each are added in quadrature to get the final systematic error. 
  $\sigma_\mathrm{stat.}$ is the statistical error determined from the DR16 data.
  }
  \begin{tabular}{cccccc}
    \hline
    \hline
    Parameter & $\sigma_\mathrm{syst,model}$ & $\sigma_\mathrm{syst,cosmo}$ & $\sigma_\mathrm{syst,tot}$ & $\sigma_\mathrm{stat}$ & $\sqrt{\sigma_\mathrm{syst,tot}^2 + \sigma_\mathrm{stat}^2}$ \\ 
    \hline
    $f\sigma_8$ & $0.0144$ &  $0.0075$ & 0.0162 & 0.077 & 0.079 \\
    $\aperp/\apar$ & $0.0042$ &  $0.0081$ & 0.0091 & 0.018 & 0.020 \\
    \hline
  \end{tabular}
  \label{table:budget}
\end{table*}

\subsection{Effect of the fiducial cosmology}
\label{sec:fiducial cosmo}
To test potential systematics associated with the arbitrary choice of the fiducial cosmology adopted for the analysis, we repeat the tests on the \nseries{} performed above using the $\Omega_m=0.310$ and $\Omega_m=0.350$ cosmological models. 
For these models, while the expected growth rate $(f\sigma_8)^\mathrm{exp}$ remains the same as before, the values of $(\aperp/\apar)^\mathrm{exp}$ are $0.99104$ and $0.97713$ for $\Omega_m=0.310$ and $\Omega_m=0.350$ respectively. 
To determine the additional contribution to the errors from the choice of cosmology, we determine the quantity $\Delta x^\mathrm{ref}\equiv||\Delta x|^{\Omega_m} - |\Delta x|^{\Omega_m^\mathrm{true}}|$, defined as the difference between the absolute values of the systematic offset in parameter $x$ when the mocks are analysed in the given cosmology and that determined in Section~\ref{sec:model_syst} above when they are analysed in the true \nseries{} cosmology, $\Omega_m^\nseries=0.286$. 
These results are summarised in Table~\ref{table:ref_cosmo_syst}. 
The measurement uncertainties for $f\sigma_8$ or $\aperp/\apar$ did not change with changes to $\Omega_m$.

The effect of the fiducial cosmology on the determination of the growth rate $f\sigma_8$ is found to be small compared to the modelling systematic determined above. 
However, for the $\Omega_m=0.350$ case a relative large shift of $0.8\%$ is seen in the recovered value of $\aperp/\apar$.

\subsection{Summary of systematic error budget}
The final systematic error budget for the measurement is summarised in Table~\ref{table:budget}. 
We separate the total systematic error into the contribution from modelling uncertainties and that from the choice of reference cosmology. 
For the modelling uncertainty, we take the larger of the offsets seen in the \ezmock{} analysis at $\Omega_m=0.310$ and the \nseries{} analysis at $\Omega_m=0.286$. 
For the error associated with the cosmological model adopted, for each parameter we take the larger of the offsets from Table~\ref{table:ref_cosmo_syst} when analysing the \nseries{} mocks at different $\Omega_m$. 
These individual contributions are added in quadrature to obtain the total systematic error budget. 
The total systematic error contribution is small compared to the statistical errors from the data and so has very little impact on the final measurements.

%%%%%% FIGURE %%%%%%
\begin{figure}
    \centering
    \includegraphics[width=\columnwidth]{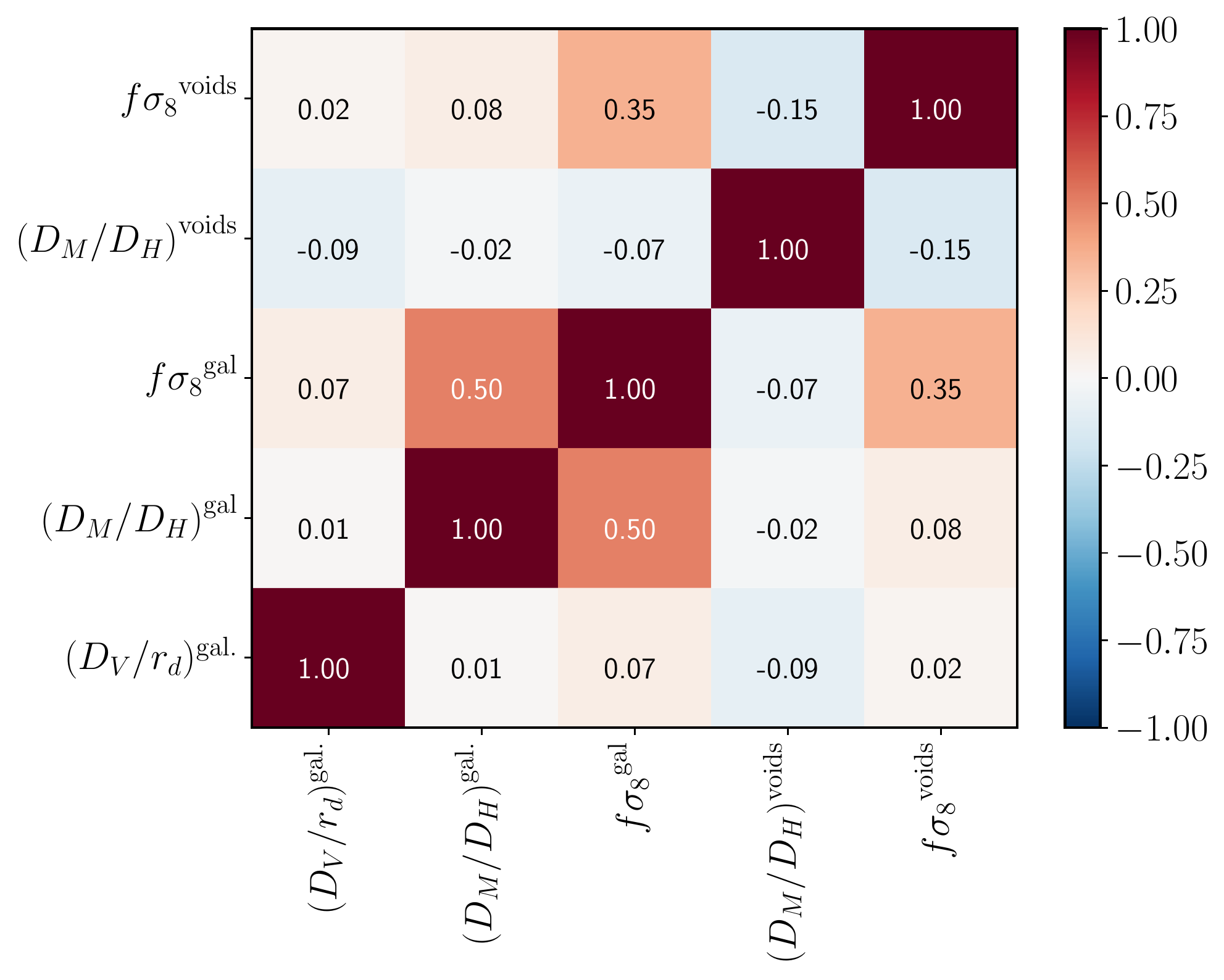}
    \caption{Correlation coefficients between the measured values of $D_V/r_d$, $D_M/D_H$ and $f\sigma_8$ obtained from the BAO+RSD consensus galaxy clustering method and the void-galaxy correlation method applied to each of the 1000 \ezmocks{}, distinguished by the superscripts $^\rmn{gal.}$ and $^\rmn{voids}$ respectively. 
    The void-galaxy method does not measure $D_V/r_d$ at all, so the corresponding entries have been omitted.
    } 
    \label{fig:cross-cov-dr16}
\end{figure}

%%%%%% FIGURE %%%%%%
\begin{figure*}
    \centering
    \includegraphics[width=0.95\columnwidth]{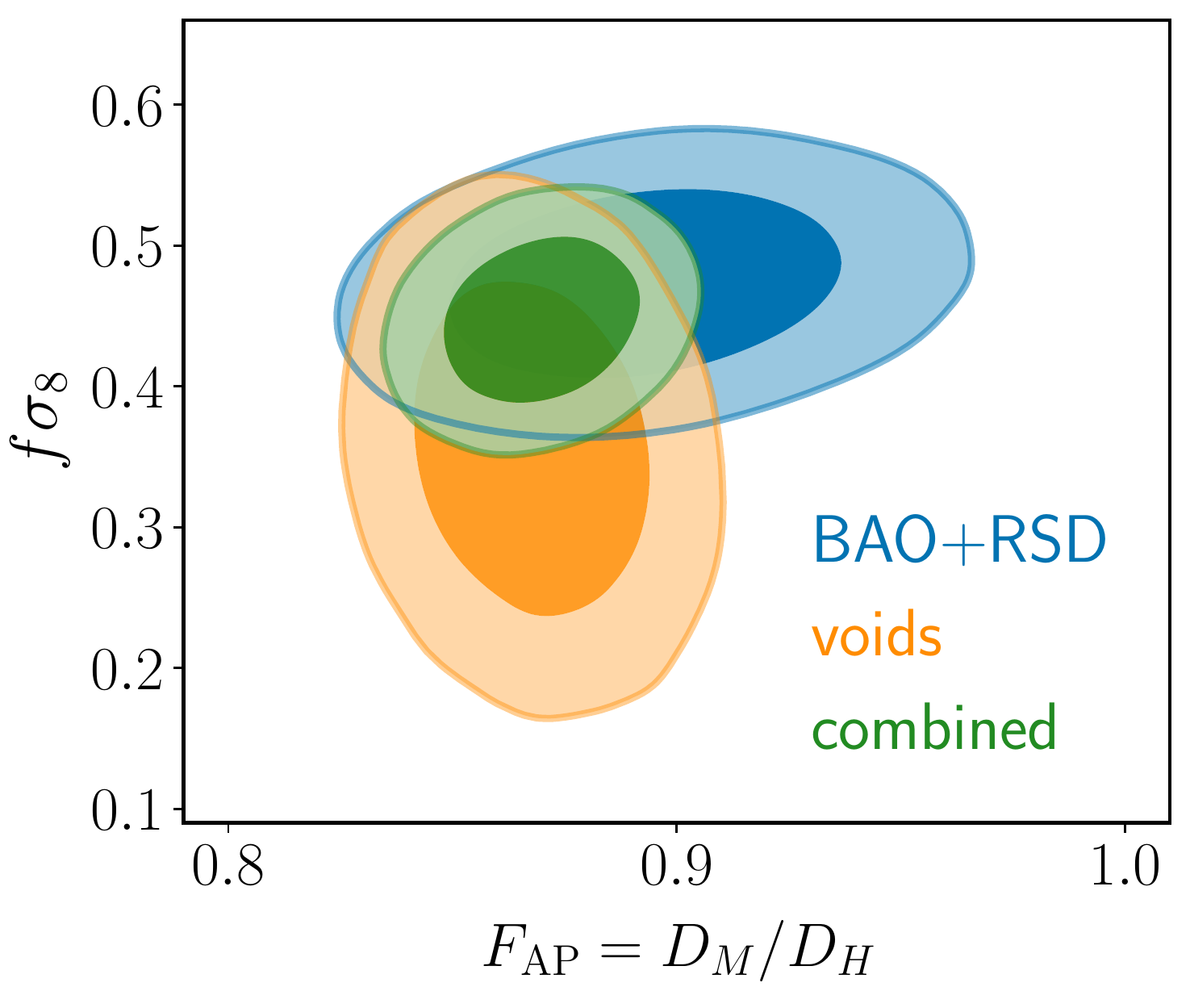}
    \includegraphics[width=0.95\columnwidth]{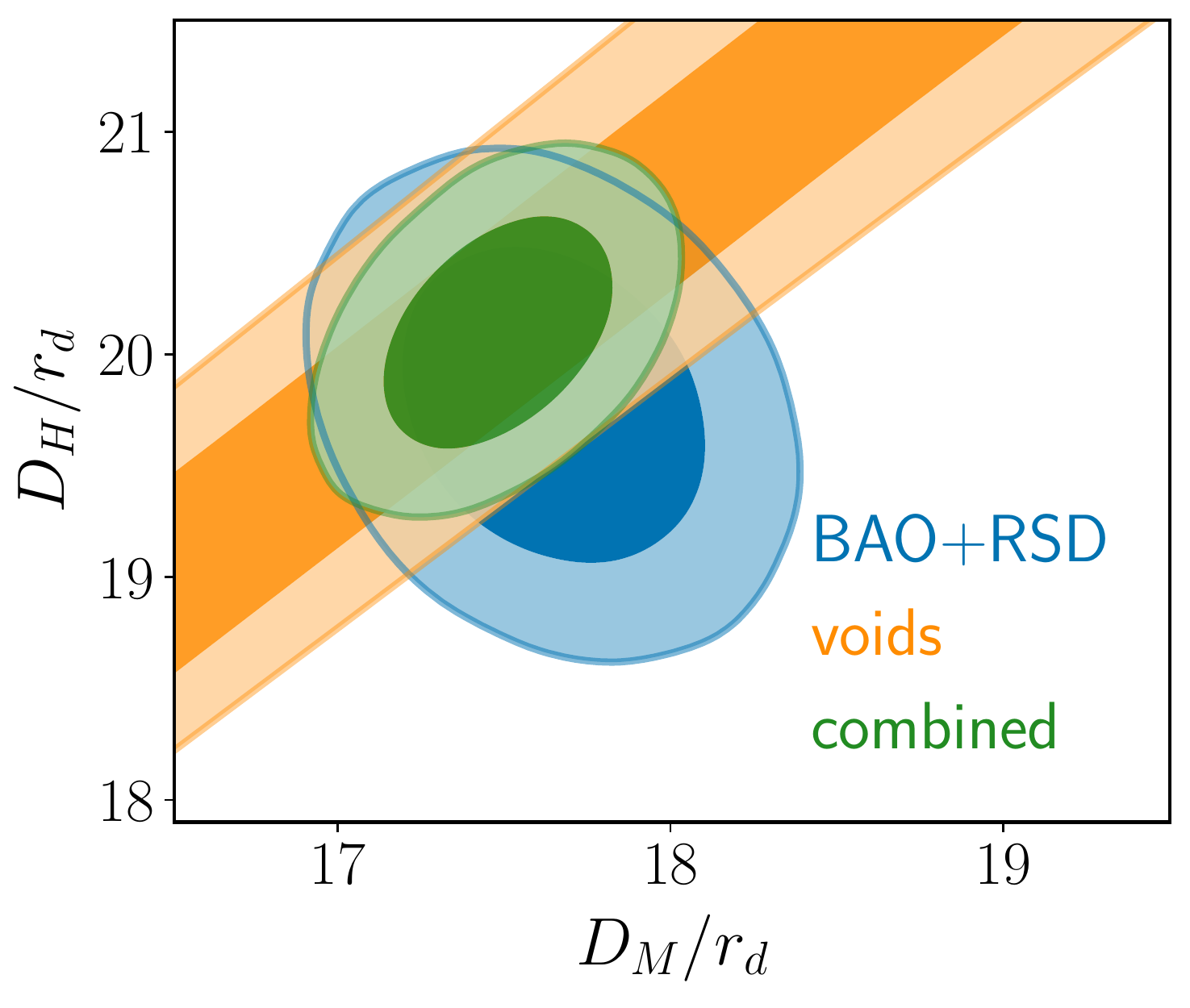}
    \caption{Marginalised $1$ and $2\sigma$ posterior constraints on cosmological parameters from the DR16 eBOSS+CMASS LRG sample at redshift $z_\rmn{eff}=0.70$ obtained from galaxy clustering (consensus of BAO+RSD; blue), void-galaxy correlation (orange) and their combination (green). 
    The left panel shows the growth rate $f\sigma_8$ versus the ratio of transverse comoving distance $D_M$ to Hubble distance $D_H$. 
    The right panel shows constraints on $D_M/r_d$ and $D_H/r_d$, where $r_d$ is the sound horizon scale. Voids provide a tight constraint on the Alcock-Paczynski parameter $D_M/D_H$. 
    All contours include the effect of systematic errors. 
    % BAO+RSD results are from \citet{LRG_corr,gil-marin20a}. 
    } 
    \label{fig:consensus1}
\end{figure*}

%%%%%% FIGURE %%%%%%
\begin{figure*}
    \centering
    \includegraphics[width=0.7\linewidth]{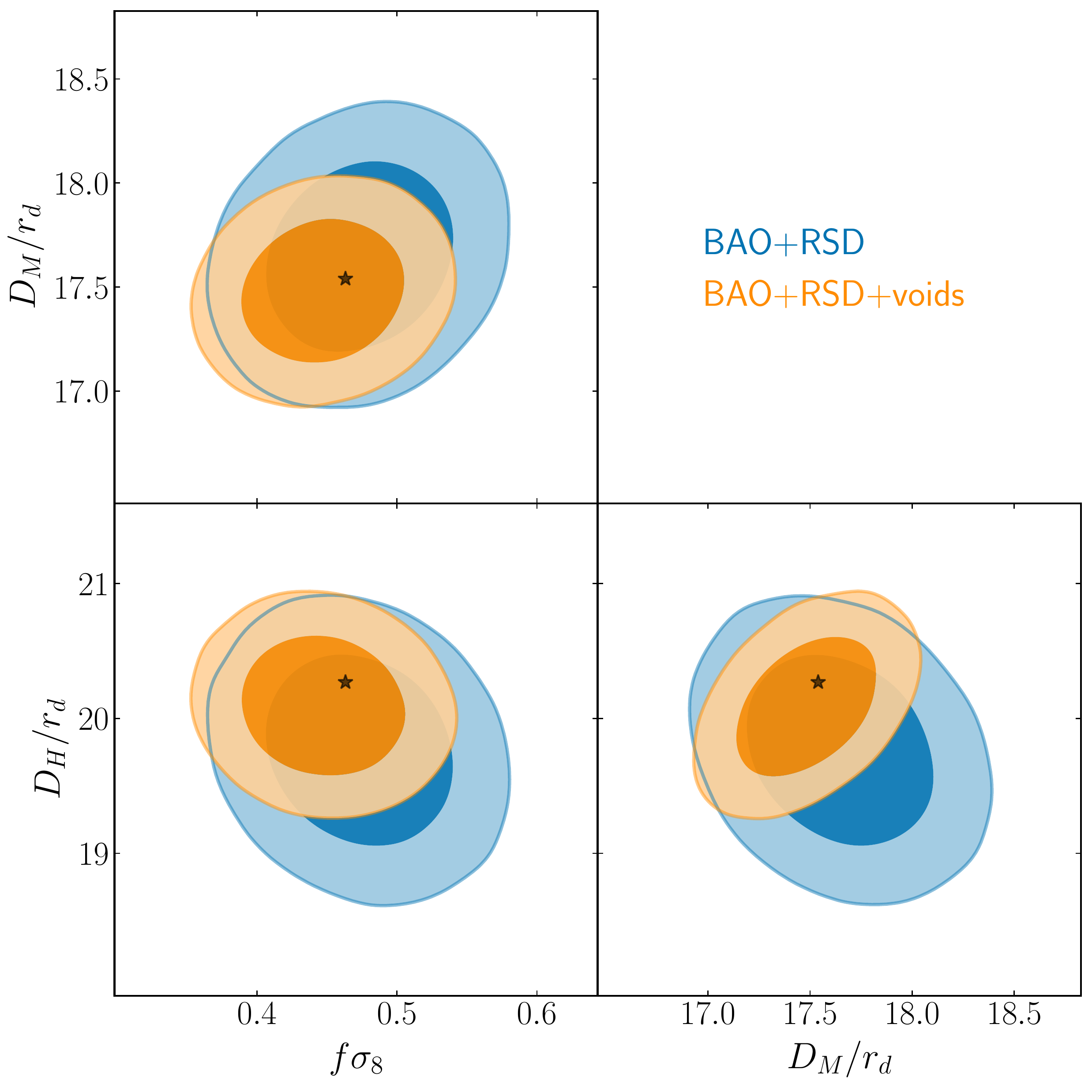}
    \caption{The final consensus measurements of $D_M/r_d$, $D_H/r_d$ and $f\sigma_8$ from the DR16 eBOSS+CMASS LRG sample using the consensus BAO+RSD galaxy clustering measurements from \citet{LRG_corr,gil-marin20a} (blue) and from the combination of these measurements with the void-galaxy results of this paper (orange). 
    The black stars in each panel indicate the expected values for a flat \lcdm{} model with parameters set to the Planck 2018 best-fit values. 
    The addition of void information reduces the uncertainty in $f\sigma_8$, $D_M/r_d$ and $D_H/r_d$ by $13\%$, $23\%$ and $28\%$ respectively. 
    } 
    \label{fig:consensus2}
\end{figure*}

%%%%%% CONSENSUS %%%%%%
\section{Consensus results}
\label{sec:consensus}
The DR16 eBOSS+CMASS LRG data analysed here has also been used to perform post-reconstruction BAO and pre-reconstruction galaxy clustering full-shape analyses in Fourier \citep{gil-marin20a} and configuration space \citep{LRG_corr}. 
While all these analyses use the same data, their information content is not the same as they include different scales in the analysis and are sensitive to different physical effects. 
This is particularly true for the void-galaxy analysis presented here, as the sharp features in the quadrupole break the degeneracy between the RSD and Alcock-Paczynski effects at scales of 20-50 $h^{-1}$Mpc whereas for galaxy clustering this is primarily achieved through observation of the BAO feature in the monopole at $\sim100\;h^{-1}$Mpc.

As a result, a very significant gain in information can be achieved by combining void-galaxy, BAO and RSD measurements to obtain a single consensus set of cosmological parameters, as first demonstrated by \citet{Nadathur:2019c, Nadathur:2020a}. 
To do so, we use the ``best linear unbiased estimator" approach for combining correlated posteriors described by \citet{Sanchez:2017a}, which has previously been used by \citet{Alam:2017}, \citet{Nadathur:2019c} and the eBOSS set of papers \citep{LRG_corr,gil-marin20a,raichoor20a,demattia20a,hou20a,neveux20a}. 
This method is based on expressing the results of each experiment performed on the same dataset in terms of a set of common cosmological parameters and building a linear estimator for the consensus values of these parameters based on the cross-covariance of the measurements determined from their application to mock galaxy samples, in this case the 1000 \ezmock{} realizations. 
In determining this consensus as described below, we will assume that the combination applies at the effective redshift $z=0.698$ determined for the galaxy clustering consensus results \citep{LRG_corr,gil-marin20a}.

The BAO and RSD methods applied to the LRG data, whether in Fourier or configuration space, measure the cosmological parameters $\left(D_M/r_d, D_H/r_d, f\sigma_8\right)$, where $r_d$ is the sound horizon scale at the drag epoch. 
In contrast, the void-galaxy method described in this paper measures only the parameters $\left(D_M/D_H, f\sigma_8\right)$. 
To consistently combine them, we therefore take the consensus BAO+RSD results computed by \citet{LRG_corr} and \citet{gil-marin20a} for each realization of the \ezmocks, and change basis in parameter space to $\left(D_V/r_d,D_M/D_H,f\sigma_8\right)$, where 
\begin{equation}
    \label{eq:DV}
    \frac{D_v}{r_d} = \left[\left(\frac{D_M}{r_d}\right)^2\frac{D_H}{r_d}\right]^{1/3}
\end{equation}
is the angle-averaged BAO distance scale.
We use the results from the two sets of methods expressed in this parameter basis from the set of 1000 \ezmocks{} to form a $6\times6$ covariance matrix $\mathbf{C}_\mathrm{tot}$, whose off-diagonal blocks describe the cross-covariance between methods.\footnote{$\mathbf{C}_\mathrm{tot}$ has dimensions $6\times6$ because for convenience we use the pre-computed BAO+RSD consensus results instead of those for the individual galaxy clustering methods. In principle any number of methods $m$ could be combined to form a $3m\times3m$ matrix.} 
The off-diagonal blocks of this covariance matrix are determined from the \ezmocks, but we replace the diagonal blocks with the values of the covariance determined from the MCMC fit to the DR16 data. 
For the void-galaxy analysis this represents a conservative choice, as the scatter in the \ezmocks{} is slightly smaller than the uncertainty in the fit to the data, but the difference is small. 
We set the $(i,j) = (4,4)$ element of $\mathbf{C}_{\mathrm{tot}}$ to be formally infinite to represent the lack of any constraint on $D_V/r_d$ from the void-galaxy method. 

With this definition of $\mathbf{C}_\mathrm{tot}$, the $3\times3$ consensus covariance matrix describing the combination of the BAO, RSD and void-galaxy methods is
\begin{equation}
    \label{eq:consensus cov}
    \mathbf{C}_\mathrm{LRG} \equiv \left(\sum_{i=1}^m \sum_{j=1}^m \mathbf{\Psi}_{ji}\right)^{-1}\,,
\end{equation}
where $\mathbf{\Psi}_{ji}$ are the block elements of the total precision matrix $\mathbf{\Psi}_\mathrm{tot}=\mathbf{C}_\mathrm{tot}^{-1}$ and $m=2$ (for combining the BAO+RSD consensus and void results). 
Then the consensus mean parameter values are
\begin{equation}
    \label{eq:consensus mean}
    \mathbf{D}_\mathrm{LRG} = \mathbf{\Psi}_\mathrm{LRG}^{-1}\sum_{i=1}^m \left(\sum_{j=1}^m \mathbf{\Psi}_{ji}\right)\mathbf{D}_i\,,
\end{equation}
with $\mathbf{\Psi}_\mathrm{LRG}=\mathbf{C}_\mathrm{LRG}^{-1}$, where $\mathbf{D}=\left(D_V/r_d,D_M/D_H, f\sigma_8\right)$ is the parameter vector. 
We then reverse the change of parameter basis to express the consensus results in terms of the conventional $D_M/r_d$, $D_H/r_d$ and $f\sigma_8$.

Fig.~\ref{fig:cross-cov-dr16} shows the correlation structure of the resultant covariance matrix $\mathbf{C}_\mathrm{tot}$. 
Consistent with the result obtained by \citet{Nadathur:2020a} for the BOSS DR12 sample, we find that the void measurement of $D_M/D_H$ is essentially independent of the measurement of any of the parameters obtained from the BAO+RSD consensus analysis of the galaxy clustering. 
We also note that the void measurements of $D_M/D_H$ and $f\sigma_8$ are weakly anti-correlated with each other, which is the opposite of the case for galaxy clustering. 
These factors explain the orientation of the constraint contours from the galaxy clustering and void analyses, which are shown in Fig.~\ref{fig:consensus1}. 
Importantly, they also mean that the information gain obtained from combining these measurements is close to optimal.

\begin{table}
  \centering
  \caption{Final results for cosmological parameters from the eBOSS+CMASS LRG sample at $z_\mathrm{eff}=0.70$. 
  Column `BAO+RSD' refers to the consensus results obtained from the combination of galaxy clustering analyses in Fourier and configuration spaces, presented in \citet{LRG_corr}; `voids' refers to the results obtained from the void-galaxy correlation presented in this work, and `BAO+RSD+voids' to the combination of these two. 
  All reported errors include systematic contributions.
  }
  \begin{tabular}{lccc}
    \hline
    \hline
    \multicolumn{1}{c}{\bf } &  \multicolumn{1}{c}{\bf BAO+RSD} &  \multicolumn{1}{c}{\bf voids} &  \multicolumn{1}{c}{\bf BAO+RSD+voids}\\
    \noalign{\vskip 3pt}\cline{2-4}\noalign{\vskip 3pt}
    Parameter &  68\% limits &  68\% limits &  68\% limits\\
    \hline
    $D_M/r_d$ & $17.65\pm 0.30$ & --- & $17.48\pm 0.23$\\
    $D_H/r_d$ & $19.77\pm 0.47$ & --- & $20.10\pm 0.34$\\
    $f\sigma_8$ & $0.473\pm 0.045$ & $0.356\pm 0.079$ & $0.447\pm 0.039$\\
    $D_M/D_H$ & $0.893\pm 0.029$ & $0.868\pm 0.017$ & $0.870\pm 0.014$\\
    $D_V/r_d$ & $18.33\pm 0.22$ & --- & $18.31\pm 0.22$\\
    \hline
  \end{tabular}
  \label{table:cosmo}
\end{table}

The final mean parameters and their $1\sigma$ errors for the BAO+RSD, voids and the full consensus results are summarised in Table~\ref{table:cosmo}. 
The void results obtained in this paper are consistent with those from BAO+RSD, but the uncertainty in measurement of the Alcock-Paczynski parameter $F_\mathrm{AP}\equiv D_M/D_H$ is reduced by a factor of over $41\%$. 
As a consequence of this and the independence of the two methods, the consensus results from the BAO+RSD+voids combination show a very significant gain in precision in the final parameters, equivalent to a reduction in the marginalised 1D errors of $13\%$, $23\%$ and $28\%$ in $f\sigma_8$, $D_M/r_d$ and $D_H/r_d$ respectively, relative to their values from BAO+RSD alone. 
This improvement in parameter constraints is shown graphically in Fig.~\ref{fig:consensus2}. 
Our final results correspond to a $1.3\%$ measurement of $D_M/r_d$, a $1.7\%$ measurement of $D_H/r_d$ and an $8.7\%$ measurement of $f\sigma_8$. 
The final consensus covariance matrix for these parameter measurements is
\begin{equation}
    \label{eq:consensus_cov}
    \mathbf{C}_{{\rm LRG}} = \begin{blockarray}{ccc}
    D_M/r_d & D_H/r_d & \fsig \vspace{1mm} \\
    \begin{block}{(ccc)}
    5.14 \times 10^{-2} & 3.22 \times 10^{-2} & 1.20 \times 10^{-3} &  \\
     -  & 1.17 \times 10^{-1} & -1.21 \times 10^{-3} &  \\
     -  &  -  & 1.51 \times 10^{-3} &  \\
    \end{block}
    \end{blockarray}.
\end{equation}
Calculating the volume of the likelihood region in parameter space as $V=(\mathrm{det}\,\mathbf{C}_\rmn{LRG})^{1/2}$, this corresponds to a $55\%$ reduction in the allowed parameter volume compared to the case for consensus BAO+RSD.

%%%%%% CONCLUSION %%%%%%
\section{Conclusion}
\label{sec:conclusions}
We have presented a cosmological analysis of the anisotropic void-galaxy correlation measured in the final DR16 eBOSS LRG sample. 
Voids were extracted using the \texttt{REVOLVER} watershed void-finder after use of a reconstruction-based RSD removal technique to remove systematic void selection bias effects and ensure validity of the modelling.
We then modelled the multipoles of the measured correlation function to perform a joint fit for the growth rate of structure and geometrical distortions due to the Alcock-Paczynski effect.
From this analysis alone we obtained constraints $f\sigma_8=0.356\pm0.079$ and $D_M/D_H=0.868\pm0.017$.
These results are in excellent agreement with flat \lcdm{} model expectations and with the corresponding results obtained from the combination of BAO and RSD in the galaxy clustering for the same sample by \citet{LRG_corr} and \cite{gil-marin20a}, but the $1.9\%$  precision of the constraint on $D_M/D_H$ from voids is more than $40\%$ better than that from the consensus galaxy clustering result.

The degeneracy directions for parameter constraints obtained from voids in this way are orthogonal to those galaxy clustering.
We quantified the cross-covariance between results from the void-galaxy analysis and those from standard galaxy clustering techniques using mocks, and showed that it is small.
As a result, the combination of our void-galaxy results with those previously obtained from BAO and full-shape RSD analyses leads to a large gain in information.
We performed this combination and obtained final consensus results $f\sigma_8=0.447\pm0.039$ ($8.7\%)$, $D_M/r_d=17.48\pm0.23$ ($1.3\%$) and $D_H/r_d=20.10\pm0.34$ ($1.7\%$) at the sample effective redshift $z_\rmn{eff}=0.70$.
These are the most precise measurements to date at this redshift: compared to the best previous results for this eBOSS sample from the consensus BAO+RSD fits \citep{LRG_corr,gil-marin20a} they represent an overall 55\% reduction in the allowed volume in parameter space for these quantities, or better than doubling the measurement precision. This gain is equivalent to more than a factor of 4 increase in the data volume of the survey compared to using galaxy clustering alone.

Our final consensus measurement of the Alcock-Paczynski parameter is $D_M/D_H=0.870\pm0.014$ at $z_\rmn{eff}=0.70$, in excellent agreement with the Planck 2018 best-fit value, extrapolated to the same redshift assuming the validity of \lcdm, $D_M/D_H=0.866\pm0.003$. 
This precision achieved in the low-redshift AP measurement strongly constrains non-\lcdm{} models with curvature or a varying dark energy equation of state \citep{Nadathur:2020a}.

The modelling and measurement methods used in this work are the same as those previously applied to the void-galaxy analysis of the BOSS LRG sample by \citet{Nadathur:2019c}. 
We have reported several tests of the robustness of this method and quantified the effects of possible systematic errors through application of the pipeline on a large number of mock galaxy catalogues generated using both full $N$-body and fast approximate methods.
In particular we tested the effects of errors due to possible inadequacy of the model and the errors due to the arbitrary choice of reference cosmology for the analysis, and included them in the final cosmological results quoted above. 
For the void analysis systematic errors increased the total error budget by $11\%$ and $2.6\%$ for $D_M/D_H$ and $f\sigma_8$ respectively.

The results in this paper are complementary to the void-galaxy analysis of \citet{Aubert20a}, which included the DR16 ELG and quasar samples in addition to the LRG sample used in this work. 
\citet{Aubert20a} used a different measurement method without the reconstruction step used here, a different model of the void-galaxy correlation, and fit for the growth rate at fixed reference cosmology without the AP distortion terms.
For the LRG sample, our results have a higher statistical precision and smaller systematic errors; we provide a comparison of the contribution to the systematic error budget from the use of different models in Appendix~\ref{sec:appendixA}. 

The considerable information gain from the void-galaxy correlation demonstrated in this work for the LRG sample highlights the potential of the voids as cosmological probes and motivates the integration of these analysis techniques into the large-scale structure toolbox of all surveys.
In the near future, DESI\footnote{\url{https://www.desi.lbl.gov/}} and Euclid\footnote{\url{https://www.euclid-ec.org/}} will probe much larger volumes of the Universe with a variety of tracers over a large range of redshifts. 
To exploit the potential of this method for these surveys it will be crucial to control systematics at an extremely low level.
The work presented here represents the most thorough investigation of these issues to date, and shows the challenge that needs to be met for future survey analyses. 

\section*{Data Availability}
The void catalogues, correlation functions, covariance matrices, and resulting likelihoods for cosmological parameters will be made available after acceptance from the \texttt{Victor} repository (\href{https://github.com/seshnadathur/victor}{https://github.com/seshnadathur/victor}) and via the SDSS Science Archive Server (\href{https://sas.sdss.org/}{https://sas.sdss.org/}) (with the exact address tbd). 
Data can also be obtained in advance of acceptance by reasonable request to the lead author.

\section*{Acknowledgements}

MA and SE acknowledge support from the French National Research Agency by the eBOSS ANR grant (ANR-16-CE31-0021) and the OCEVU Labex (ANR-11-LABX-0060).
GR acknowledges support from the National Research Foundation of Korea (NRF) through Grants No. 2017R1E1A1A01077508 and No. 2020R1A2C1005655 funded by the Korean Ministry of Education, Science and Technology (MoEST), and from the faculty research fund of Sejong University. 

Funding for the Sloan Digital Sky Survey IV has been provided by the Alfred P. Sloan Foundation, the U.S. Department of Energy Office of Science, and the Participating Institutions. 
SDSS-IV acknowledges support and resources from the Center for High-Performance Computing at the University of Utah. 
The SDSS web site is www.sdss.org.

SDSS-IV is managed by the Astrophysical Research Consortium for the 
Participating Institutions of the SDSS Collaboration including the 
Brazilian Participation Group, the Carnegie Institution for Science, 
Carnegie Mellon University, the Chilean Participation Group, the French Participation Group, Harvard-Smithsonian Center for Astrophysics, 
Instituto de Astrof\'isica de Canarias, The Johns Hopkins University, Kavli Institute for the Physics and Mathematics of the Universe (IPMU) / 
University of Tokyo, the Korean Participation Group, Lawrence Berkeley National Laboratory, 
Leibniz Institut f\"ur Astrophysik Potsdam (AIP),  
Max-Planck-Institut f\"ur Astronomie (MPIA Heidelberg), 
Max-Planck-Institut f\"ur Astrophysik (MPA Garching), 
Max-Planck-Institut f\"ur Extraterrestrische Physik (MPE), 
National Astronomical Observatories of China, New Mexico State University, 
New York University, University of Notre Dame, 
Observat\'ario Nacional / MCTI, The Ohio State University, 
Pennsylvania State University, Shanghai Astronomical Observatory, 
United Kingdom Participation Group,
Universidad Nacional Aut\'onoma de M\'exico, University of Arizona, 
University of Colorado Boulder, University of Oxford, University of Portsmouth, 
University of Utah, University of Virginia, University of Washington, University of Wisconsin, 
Vanderbilt University, and Yale University.

%%%%%%%%%%%%%%%%%%%%%%%%%%%%%%%%%%%%%%%%%%%%%%%%%%



\bibliographystyle{mnras}
\bibliography{references} 

%%%%%%%%%%%%%%%%%%%%%%%%%%%%%%%%%%%%%%%%%%%%%%%%%%

%%%%%%%%%%%%%%%%% APPENDICES %%%%%%%%%%%%%%%%%%%%%

\appendix

\section{Comparison of systematics in alternative models of the void-galaxy correlation}
\label{sec:appendixA}

%%%%%% FIGURE %%%%%%
\begin{figure*}
    \centering
    \includegraphics[width=0.8\linewidth]{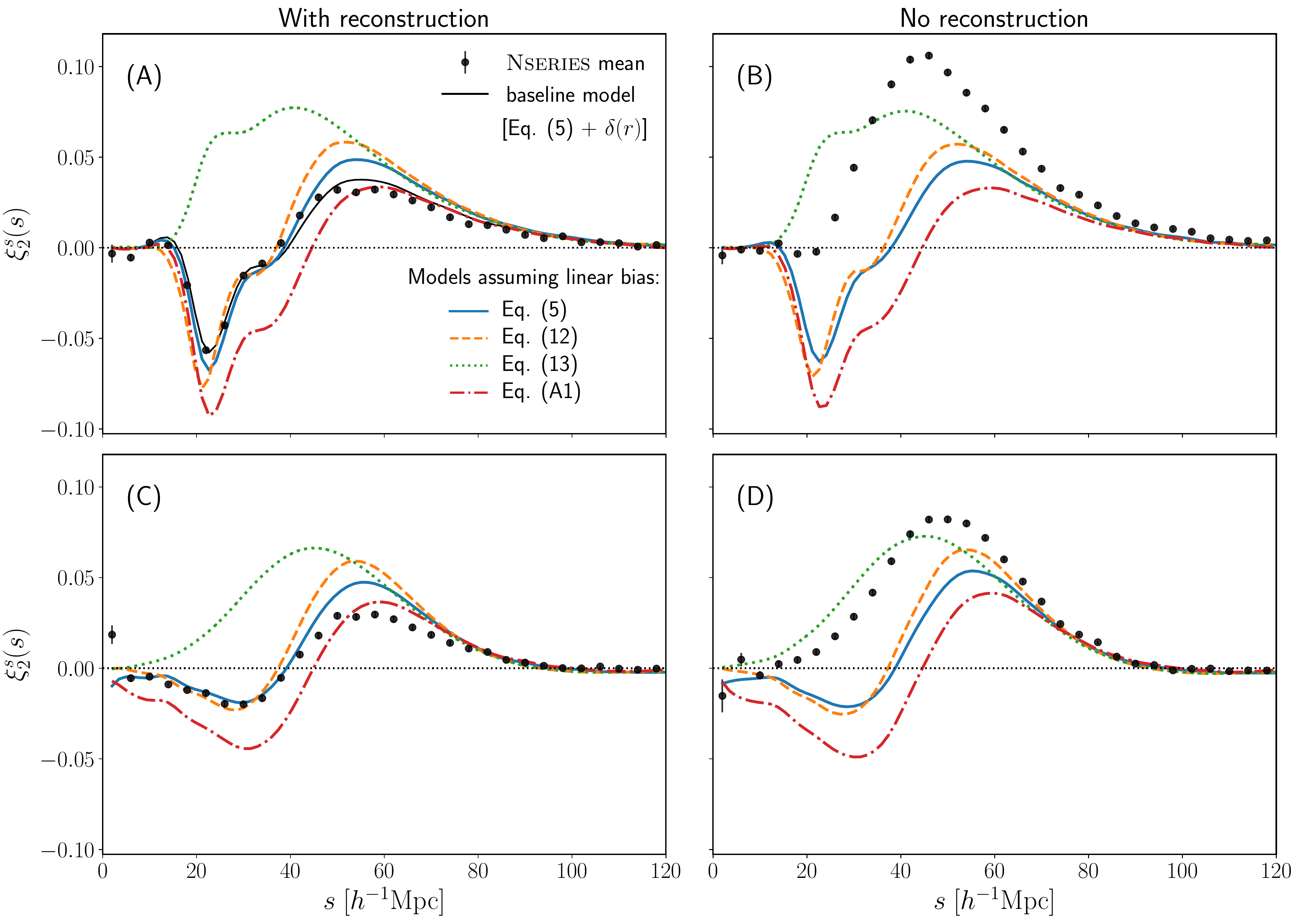}
    \caption{Comparison of the mean quadrupole moment $\xi^s_2(s)$ measured in the \nseries{} mocks and the predictions of several alternative models described in Appendix~\ref{sec:appendixA} and the listed equations. 
    The left-hand panels (A) and (C) show the case where voids are identified in the RSD-removed galaxy field obtained from the reconstruction step; for panels (B) and (D) this reconstruction step is omitted.
    Panels (A) and (B) are for void centres defined as the points of minimum density, as used in the main analysis throughout the rest of the paper.
    Panels (C) and (D) are for an alternative centre definition based on the barycentre of galaxy positions.
    All models were calculated using the open-source \texttt{Victor} code, using the same input $\xi^r(r)$ function in each panel, evaluated at the same fiducial value of $\beta=f/b=0.40$, and using the linear bias approximation $\delta(r)=\xi^r(r)/b$ as described in the text.
    In panel (A) for comparison we additionally show the model of Eq.~\ref{eq:full_model} without the linear bias assumption, as used in our main analysis.
    Data points in panels (A) and (C) correspond to reconstruction performed with $\beta=0.40$.
    Error bars on the data points in each panel correspond to the error in the mean over the \nseries{} mocks and are therefore a factor of $\sqrt{84}$ smaller than the errors on an individual mock.
    The same void selection cut $R_v>49\;h^{-1}$Mpc is used in all cases.
    } 
    \label{fig:model_comp}
\end{figure*}

\begin{table*}
  \centering
  \caption{A comparison of systematic errors in parameter inference from two alternative models of the void-galaxy correlation. 
  Eq.~\ref{eq:multipole ratio} is the multipole ratio estimator for $\beta=f/b$ derived from the model of Eq.~\ref{eq:linear_approx}, but does not include Alcock-Paczynski distortions. 
  Eq.~\ref{eq:GSM} describes the Gaussian streaming model, which has the same parameters as for our baseline model. 
%   Eq.~\ref{eq:GSM} is applied to the same void sample as used in our baseline analysis, including reconstruction. 
  Details of the application of each model are provided in the text; the same void selection cut $R_v>49\;h^{-1}$Mpc is used in all cases.
  Modelling errors are calculated in the same way as for Table~\ref{table:model_err}. 
  The uncertainties in the recovered parameters (reported at the $2\sigma$ level) refer to the mean of the mocks, so are related to the uncertainty on an individual mock by a factor of $1/\sqrt{N_\mathrm{mocks}}$. 
  In calculating $\beta^\mathrm{exp}$ we have used the fiducial values $b=2.30$ for the \ezmocks{} and $b=1.9$ for the \nseries.
  }
  \begin{tabular}{cccccccccc}
    \hline
    \hline
    Model & Mock & $N_\mathrm{mocks}$ & ref. cosmology & $\beta^\mathrm{exp}$ & $\Delta\beta\pm2\sigma$ & $(f\sigma_8)^\mathrm{exp}$ & $\Delta (f\sigma_8)\pm2\sigma$ & $(\aperp/\apar)^\mathrm{exp}$ & $\Delta(\aperp/\apar)\pm2\sigma$\\ 
    \hline
    \multirow{2}{*}{Eq.~\ref{eq:multipole ratio}} & \ezmocks & 1000 & $\Omega_m=0.310$ & 0.3527 & $-0.031\pm0.004$ & 0.4687 & --- & 0.9987 & --- \\
    & \nseries & 84 & $\Omega_m=0.286$ & 0.3979 & $-0.026\pm0.012$ & 0.4703 & --- & 1.0000 & --- \\
    \hline
    \multirow{2}{*}{Eq.~\ref{eq:GSM}} & \ezmocks & 1000 & $\Omega_m=0.310$ & 0.3527 & --- & 0.4687 & $-0.046\pm0.026$ & 0.9987 & $-0.0241\pm0.0049$ \\
    & \nseries & 84 & $\Omega_m=0.286$ & 0.3979 & --- & 0.4703 & $-0.040\pm0.026$ & 1.0000 & $-0.0595\pm0.0060$ \\
    \hline
  \end{tabular}
  \label{table:alt_model_err}
\end{table*}

As discussed in Section~\ref{sec:model}, several different models of the void-galaxy correlation have been used in the literature. 
We have provided a thorough investigation of the modelling systematics associated with the linear dispersion model of Eq.~\ref{eq:full_model} used in this work and demonstrated its robustness. 
Our aim here is to provide a comparison between this model and other alternatives.

To provide the reader a feel for the differences between models that are currently used, in Fig.~\ref{fig:model_comp} we plot the predictions for the void-galaxy quadrupole moment $\xi^s_2(s)$ compared to the mean quadrupole measured in the \nseries{} mocks.
The first alternative model we plot is the Kaiser model analogue \citep{Cai:2016a} described by Eq.~\ref{eq:Kaiser limit}. 
This corresponds to the limit of our baseline linear dispersion model, Eq.~\ref{eq:full_model}, in the limit of negligible velocity dispersion. 
The second alternative model is that of Eq.~\ref{eq:linear_approx}, which is an approximation to Eq.~\ref{eq:Kaiser limit} and is the model that has been used by \citet{Hamaus:2017a,Hawken:2020,Achitouv:2019,Aubert20a}. 
Finally, we include the model of \citet{Paz:2013,Cai:2016a}, which is often referred to as the ``Gaussian streaming model" \citep{Hamaus:2016,Hawken:2017,Achitouv:2017a}:
\begin{multline}
    \label{eq:GSM}
    1 + \xi^s(s,\mu) = \int\left[1+ \xi^r\left(r\right)\right] \times \\ \frac{1}{\sqrt{2\pi}\sigma_{v_\parallel}(r)} \exp\left(-\frac{\left(v_\parallel-v_r(r)\mu\right)^2}{2\sigma_{v_\parallel}^2(r)} \right)\,dv_{||}\,,
\end{multline}
where $r\equiv\sqrt{r_\perp^2+r_\parallel^2}$, with $r_\perp=s_\perp$ and $r_\parallel=s_\parallel-v_\parallel/aH$. 
All of these models are implemented in the public code \texttt{Victor} as options that can be selected by the user.

In order to provide full and fair comparison, we use the same input function $\xi^r(r)$ for each of the models, which is calculated using void catalogues from the 84 \nseries{} mocks, analysed in the $\Omega_m=0.286$ model, and determined from the cross-correlation of these voids with the respective RSD-removed galaxy fields, obtained in each case using the reconstruction procedure with the fiducial value $\beta=f/b=0.40$ for \nseries.
For comparison, we show the mean quadrupole moment $\xi^s_2(s)$ of the cross-correlation of the voids with the original redshift-space galaxy field as the black data points in the figure (with error bars corresponding to the error in the mean over the 84 mocks).
To cover all permutations, Figure \ref{fig:model_comp} is shown with 4 panels, labelled (A)-(D).
The left panels correspond to the case where voids are identified in the post-reconstruction, RSD-removed galaxy field as in the primary analysis throughout the paper; for the right panels no reconstruction was applied, matching the practice in all papers where the models of Eqs.~\ref{eq:linear_approx} and \ref{eq:GSM} have previously been applied to data.
The top panels correspond to cross-correlations measured using void centres located at the minimum density point, which is the default used throughout the paper.
The bottom panels use an alternative void centre definition corresponding to the Voronoi-volume-weighted barycentre of galaxy positions within the void, matching the practice in \citet{Hamaus:2017a,Hawken:2020,Aubert20a}.
In each panel we also show the model predictions for the models of Eqs.~\ref{eq:full_model}, \ref{eq:Kaiser limit}, \ref{eq:linear_approx} and \ref{eq:GSM}.
To treat them all on the same footing, we use the approximation $\delta(r)=\xi^r(r)/b$ for the void matter density profile in each case, where $b=1.9$ is the effective linear galaxy bias value for the \nseries{} mocks, and all models were calculated at the fiducial growth rate $f=\beta b=0.756$ for the \nseries{} mocks at $z=0.55$.
In addition, for the models of Eq.~\ref{eq:full_model} and Eq.~\ref{eq:GSM} we use the same velocity dispersion relation $\sigma_{v_\parallel}(r)$ described in Section~\ref{sec:model}, with $\sigma_v=380\;\mathrm{km\,s}^{-1}$.
The linear bias assumption here is different to our main analysis, which uses the template $\delta(r)$ profile calibrated from the \bigmd{} mock. 
To indicate the effect of this in panel (A) we include the model predictions with the calibrated $\delta(r)$ for comparison.
For panels (A) and (C), the \nseries{} data are shown for the same fiducial value $\beta=0.40$, while for the other two panels the data vector is independent of $\beta$.
It should be emphasised that none of the model curves plotted in this figure are fits to the data: they are all shown at the fiducial values described above.
For conciseness and clarity of the figure, we do not show the monopole moments for any of these scenarios.
However for quantitative results both monopole and quadrupole moments would need to be fitted.

Figure \ref{fig:model_comp} highlights several important general aspects of the void-galaxy correlation.
Firstly, the measured data vector---the quadrupole moment $\xi^s_2(s)$ of which is shown here---changes substantially depending on whether voids are identified in the RSD-removed galaxy field or directly in redshift-space.
This is due to the effects of the RSD mapping on the set of void centre positions described by \citet{Nadathur:2019b}.
It can alternatively (and perhaps more intuitively) be viewed as a strong selection bias effect: the operation of void-finders on the  redshift-space galaxy field breaks the assumed real-space isotropy by preferentially selecting voids with a stronger velocity outflow along the line-of-sight direction (see Figure 2 of \citealt{Nadathur:2019b}).
For voids found in the RSD-removed galaxy field (panels (A) and (C)), the measured quadrupole generically shows a negative dip followed by a zero-crossing.
This qualitative feature is shared by the models of Eqs.~\ref{eq:full_model}, \ref{eq:Kaiser limit} and \ref{eq:GSM}, but not by Eq.~\ref{eq:linear_approx}, even though it is intended as an approximation to Eq.~\ref{eq:Kaiser limit}.\footnote{The difference for Eq.~\ref{eq:linear_approx} is primarily due to the additional approximation $s\simeq r$ commonly used in the literature. If the correct expression for $s$ is used, Eq.~\ref{eq:linear_approx} \emph{also} generically predicts a negative dip and zero-crossing of the quadrupole, though still differing from Eq.~\ref{eq:Kaiser limit}.} However, Eq.~\ref{eq:GSM} does not reduce to the Kaiser limit form of Eq.~\ref{eq:Kaiser limit} in the limit $\sigma_{v_\parallel}\rightarrow0$.

The second obvious feature is that the data vector also depends strongly on the choice of void centre definition, even when the population of voids used is identical (compare panel (A) to panel (C), or panel (B) to panel (D)).
The model predictions for different centres \emph{also} differ, because the change in the void centre definition also affects the measured monopole moments, and thus the input $\xi^r(r)$ to the theory calculation.
However, for a fixed void centre choice, the use or not of the RSD-removal step prior to void finding has relatively little effect on the real-space monopole $\xi^r(r)$ (although strongly affecting the quadrupole), and so the model predictions do not change much across a row.

It is important to note here that \emph{all} of the models shown here are derived from the same fundamental assumptions, which are only satisfied with the use of reconstruction \citep{Nadathur:2019b}, i.e. in panels (A) and (C).
Irrespective of the centre choice, in both of these panels Eq. \ref{eq:full_model} provides a good description of the \nseries{} data, which is further improved when the linear bias assumption imposed here is dropped, and Eqs.~\ref{eq:Kaiser limit} and \ref{eq:GSM} capture the qualitative features of the data, although with worse quantitative fits.
When reconstruction is not applied prior to void-finding, the situation is unsurprisingly different.
In the special case of using the void barycentre definition (panel (D)), Eq.~\ref{eq:linear_approx} provides a better qualitative description of the observed quadrupole as shown, although not of the monopole, and with a poor $\chi^2$ \citep[see also the model comparisons in][]{Nadathur:2019a}.
However this qualitative agreement is coincidental and not robust against changes in the void centre definition, as seen in panel (B).

While these observations are strongly suggestive, all models are ultimately only approximations to the truth. 
Therefore the real question is which of these models are \emph{useful}, to be judged in terms of whether they allow unbiased recovery of the cosmological parameters of interest. 
To test this, we repeat the tests of modelling systematics described in Section~\ref{sec:model_syst} for the models of Eq.~\ref{eq:linear_approx} and Eq.~\ref{eq:GSM}. 
That is, we test how accurately they recover the fiducial cosmological parameters in the \ezmocks{} and \nseries{} mocks when analysed in the cosmological models with $\Omega_m=0.310$ and $\Omega_m=0.286$ respectively. 

We wish to compare compare the performance of each model in the particular circumstance in which it performs best. 
For the model of Eq.~\ref{eq:linear_approx} it is clear from Figure~\ref{fig:model_comp} that this corresponds to that shown in panel (D), i.e. without prior reconstruction and using barycentres.
This also matches the scenario under which this model has previously been used in the literature.
We recompute the covariance matrix to match these choices.
A full application of Eq.~\ref{eq:linear_approx} requires knowledge of the real-space correlation monopole $\xi^r(r)$, which could be estimated from correlation with the RSD-removed galaxy field in the same way as for our main analysis.
However, as might be guessed from Figure~\ref{fig:model_comp}, we found this generally led to poor $\chi^2$ and highly biased parameter estimates \citep[see also][]{Nadathur:2019a}, and the model cannot successfully reproduce both multipoles $\xi^s_0(s)$ and $\xi^s_2(s)$ simultaneously.
In most works in the literature, the model of Eq.~\ref{eq:linear_approx} is instead used in the form of a `multipole ratio' estimator (\citealt{Cai:2016a}; see also \citealt{Hamaus:2017a,Achitouv:2019,Hawken:2020,Aubert20a}), which relates the observed monopole and quadrupole moments to the growth parameter $\beta$ by
\begin{equation}
    \label{eq:multipole ratio}
    \frac{\xi^s_2(s)}{\xi^s_0(s) - \overline{\xi^s_0}(s)} = \frac{2\beta}{3+\beta}\,,
\end{equation}
with $\overline{\xi^s_0}(s)\equiv 3/s^3\int_0^s\xi^s_0(y)y^2\,\rmn{d}y$.
This estimator performs better and does not require knowledge of $\xi^r(r)$, albeit at the expense of fixing the cosmological model and not including Alcock-Paczynski distortions.
We use this estimator for $\beta$ in our test of the model, and refer to this as model~\ref{eq:multipole ratio}.

As Eq.~\ref{eq:GSM} predicts the same negative feature in the quadrupole as our baseline model Eq.~\ref{eq:full_model}, it performs better when applied to void catalogues obtained after the reconstruction step.
As seen from Figure~\ref{fig:model_comp}, in this scenario the choice of void centre does not strongly affect the quality of the fit, although detectable features in the signal are suppressed when using barycentres.
We therefore test this model using the same baseline scenario as for the main analysis of this paper, corresponding to panel (A) of Figure~\ref{fig:model_comp}.
In particular, we use the same data vector, covariance, the same calibrated functions $\delta(r)$ and $\sigma_{v_\parallel}(r)$, and fit to the same set of free parameters as for our main analysis. 

The results of these tests are summarized in Table~\ref{table:alt_model_err}. 
For the multipole ratio model we find a statistically significant systematic offset in the recovered $\beta$, at the level of $|\Delta\beta|=0.031$. 
This agrees well with the systematic error estimated by \citet{Aubert20a}, who found $\sigma_\mathrm{syst}^\beta=0.037$ from modelling errors alone, though our error is slightly smaller. 
The small difference is due to the slightly different method used by \citet{Aubert20a}, who apply different void sample selection cuts and rescale all void-galaxy pair separations by the void radius when measuring the correlation function. 
While this systematic offset is relatively small compared to the statistical uncertainty $\sigma_\rmn{stat}^\beta=0.075$ in the DR16 eBOSS+CMASS data reported by \citet{Aubert20a}, using $b\sigma_8=1.20\pm0.05$ \citep{LRG_corr,gil-marin20a} it translates to $\sigma_\mathrm{syst}^{f\sigma_8}=0.041$, which is more than twice as large as the total systematic error we report in Table~\ref{table:budget}.

The Gaussian streaming model of Eq.~\ref{eq:GSM} performs less well, with large systematic errors in both parameters, $\sigma_\mathrm{syst}^{f\sigma_8}=0.046$ and $\sigma_\mathrm{syst}^{\aperp/\apar}=0.06$. 
This is because while this model shows a similar sign change in the quadrupole to that of our baseline model, it provides a worse quantitative fit to the data \citep{Nadathur:2019a}. 
We note that the large systematic offsets we see for this case are in qualitative agreement with those of \citet{Hamaus:2015}, who also reported systematic errors in excess of three times the statistical uncertainties for this model.

%%%%%%%%%%%%%%%%%%%%%%%%%%%%%%%%%%%%%%%%%%%%%%%%%%

% Don't change these lines
\bsp	% typesetting comment
\label{lastpage}
\end{document}